## PART I: THE MAXWELLIAN

# A THEORETICAL INVESTIGATION OF THE MAXWELLIAN VELOCITY DISTRIBUTION FUNCTION WITH CONNECTION TO CONTINUUM TRANSPORT PHENOMENA

Charles Cook
Department of Engineering Mechanics, Virginia Tech
Blacksburg, VA 24061
June 2021

ABSTRACT

The Euler and Navier-Stokes fluid mechanics equations are derived using a modified statistical mechanical approach using theory taken from the Chapman-Enskog perturbation analysis used to support the lattice Boltzmann method. Additional distributions such as the velocity vector and total scalar energy distributions are established. A complete discretization in velocity space is provided. Benchmark problems are established for simple cases modeling isothermal compressible inviscid flows. Thermal viscous flows will be the focus of future work. Overall, a more fundamental description of mass, momentum, and energy transport is uncovered and provides insights into the mathematical nature of the continuum transport equations such as the incorporation of viscosity and thermal conductivity into space and time dependence.

## INTRODUCTION

A modified statistical mechanical derivation of the compressible Navier-Stokes-Fourier equations will be established using an alternate but similar method in comparison with the commonly used Chapman-Enskog multiscale expansion [7]. A more direct approach is established leaving less room for interpretation. However, both methods still lack an appropriate numerical scheme that follows both multiscale methods. The newly modified method does not require the use of a total distribution function in reference to the full-form integro-differential Boltzmann equation. This leaves only the equilibrium distribution and its perturbations in the evolution equations. Lattice Boltzmann relaxation times are eliminated. There is only one finite time interval which comes from the truncated Taylor series expansion which has dissipation properties to be derived in future work.

A modified total energy distribution function is established and is an extension to the one provided in [12]. This is necessary due to the newly derived method as summarized above. The additional distribution is used to fill in missing transport information from the use of a single velocity distribution function. The added components are a bulk viscosity coefficient in the momentum equation, an adjustable specific heat ratio, and an adjustable spacetime dependent Prandtl number. Overall, a more fundamental description of mass, momentum, and energy transport is reviewed and expanded upon. The method provided incorporates the coefficients of viscosity and thermal conductivity into space and time dependence. Empirical studies are still required, however, the method shows possible deeper interpretations of the nature of these coefficients.

A modified Gauss-Hermite weight and lattice derivation procedure will be shown how to derive full and complete random velocity lattices which recover the desired moments exactly without error. Examples of shortened approximated lattices based on the work of [8,9] are also provided for comparison. It is however found that to a degree, more lattices lead to smoother numerical solutions. Even though the errors in the simplified lattices are approximately of order $10^{-15}$, an exact discretization exists, is simpler to derive, and may benefit stability in numerical analysis. A complete list of constraints is provided to derive the necessary Gauss-Hermite weights and associated lattices which is an extension to the one given in [7]. Within the Navier-Stokes framework, assuming a linearly viscous fluid, minimum assumptions are made to derive the momentum and

email: cook9@vt.edu https://github.com/cook9/Maxwell-Boltzmann

energy equations. The associated velocity discretizations are given without simplifications meaning the fluid or gas at hand is considered compressible, viscous, thermal, and unsteady thus giving a general mathematical description. Approximations and simplifications can be made from the model if desired without additional derivation.

The Navier-Stokes equations are a mathematical system of coupled partial differential equations containing multiple unknown variables, i.e., density, velocity, and pressure. The discretization process in space and time for numerical analysis is a complicated task due to the complex coupling of these equations. To apply techniques such as finite difference or finite volume methods, to not only solve the N-S equations efficiently and accurately for common physical phenomena, but to model extreme environment scenarios such as shock wave formation during high-speed flight, a reorganization and more fundamental description is desired. It will be described in detail how the information contained in the N-S transport equations can be stored and extracted using velocity distribution functions. The hypothesis is that rather than directly solving the system of PDE's for each spacetime variable, the transport formulas can be written in terms of a single or multiple distribution functions. In other words, distribution may be simulated over space and time which is directly connected to the three summational invariants (mass, momentum, and energy) through its moments or expectation values.

The field of computational fluid dynamics can be separated into three general categories depending on level of physical description. One, as recently explained, are conventional finite difference and finite volume techniques which directly solve the transport equations at the system or "macroscopic" level. At the other end of the spectrum are molecular simulations. The direct simulation Monte Carlo method is an in-between method which leans in the direction of molecular encounters. These types of methods aim to describe a system based on the behavior of its individual molecules. In the middle lies the lattice Boltzmann method which is set up to operate as a distribution solver from which the system information can be extracted, which is the foundation for this work. The focus will veer from traditional LBM mathematical and implementation methods. This work is aimed to establish distribution evolution equations from which the N-S equations can be directly derived and eventually be used as a replacement for which finite difference, volume, and/or element methods can be applied.

There are several reasons why current traditional lattice Boltzmann methods require more theoretical work. For one, the use of the "streaming-collision" algorithm and general boundary conditions lack a rigorous mathematical foundation. It is common for the lattices to be interpreted as particles but it will be shown in this work how they can be connected to the spatial grid. Distributions are broken up into parts then streamed to different lattice locations based on simple finite difference schemes applied in space and time. Upon completion, the updated moments are generated based on the new discrete distribution values at each spatial node. In addition, the LBM theory is based on the Chapman-Enskog multiscale analysis which differs from the original [1,17] where actual perturbations of the total distribution function are calculated from which transport equations can be derived. For example, beyond equilibrium is $\mathbb{f}^{(\mathbb{1})}$ which is in terms of $\mathbb{f}^{(\mathbb{0})}$ along with velocity and temperature gradients. The LBM framework does not calculate the perturbed distribution but uses an ad hoc combination of Taylor series and multiscale perturbation analysis to show similarities between the lattice Boltzmann equation and the Navier-Stokes equations. There is not a direct connection therefore the correlation between simulation and theory may differ. In other words, calculations made in the LBE cannot be directly mapped back to the N-S equations. Lastly and oddly, the LBM is commonly used to simulate fluids governed by the assumption that compressibility is negligible. Therefore, the general focus is mainly on the incompressible N-S momentum equation. However, the main algorithm is to update the density and velocity per time step at each grid point which is in contrast to the incompressible assumption [4]. This is a result due to the lack of connection between the LBE and transport equations.

A major focus of this work includes the use of distribution functions which are taken from the realm of probability theory. So, they are functions of a random variable which adds another unknown quantity into the analysis. The extraction of system variables from moment or expectation operations will be utilized. For numerical analysis, it is optimal to convert these integrals into finite summations. This will take the continuous random variable and convert it to a predetermined finite set of numbers which will remain independent from time and space using a certain quadrature transformation. The distribution functions that will be utilized should be thought of more in the sense of functions containing information which can be extracted upon certain mathematical operations. A major objective is to reform the N-S equations into a more manageable set of PDE's by reforming in terms of these collective distributions. The layout of this paper is as follows:

- The Maxwellian will be derived along with associated moments.
- The theoretical spacetime dependent distribution will be expanded using a Taylor series to derive the Navier-Stokes equations governing a monatomic ideal gas.
- A total energy distribution will be established and the N-S equations rederived for polyatomic gases using a modified double distribution function approach.
- Both the Maxwellian and total energy distributions will be expanded using the Hermite series then discretized in random velocity space using a Gauss-Hermite quadrature transformation.
- The discrete Maxwellian will be analyzed through the use of the Fourier series.
- A simple numerical scheme in space and time will be applied to some benchmark problems to show validity of the method.

It will be noted that the Einstein summation notation is adopted unless otherwise specified. Associated MATLAB and Maple codes can be found at the link located to the bottom right corner of the first page.

## NOMENCLATURE

| | |
|---|---|
| $D$ | number of Euclidean or spatial dimensions |
| $\rho$ | mass density, $\mathrm{kg/m^3}$ |
| $\mathbf{v}$ | velocity, $\mathrm{m/s}$ |
| $\boldsymbol{v}$ | peculiar or relative velocity, $\mathrm{m/s}$ |
| $\boldsymbol{p}$ | momentum, $\mathrm{kg \cdot m/s}$ |
| $T$ | temperature, K |
| $P$ | pressure, $\mathrm{kg/(m \cdot s^2)}$ |
| $E$ | specific energy, $\mathrm{m^2/s^2}$ |
| $\rho\mathbf{v}$ | momentum density $\mathrm{kg/(m^2 \cdot s)}$ |
| $\rho E$ | energy density, $\mathrm{kg/(m \cdot s^2)}$ |
| $\theta, \vartheta^2$ | specific thermal energy, $\mathrm{m^2/s^2}$ |
| $m$ | mass per object, kg |
| $n$ | number of objects (particles, molecules, atoms, etc.) |
| $V$ | unit volume, $\mathrm{m^3}$ |
| $\boldsymbol{\xi}$ | random variable or random velocity, $\mathrm{m/s}$ |
| $\mathrm{d}\boldsymbol{\xi} = \prod_{i=1}^{D} \mathrm{d}\xi_i$ | random velocity differential, $(\mathrm{m/s})^D$ |
| $\boldsymbol{\chi}$ | random variable or random distance, m |
| $\tau$ | time interval |
| $\Delta t$ | dissipation time |
| $k, k_B$ | Boltzmann's constant, $\mathrm{kg/K \cdot m^2/s^2}$ |
| $R, R_s$ | specific gas constant, $\mathrm{1/K \cdot m^2/s^2}$ |
| $\mathfrak{R}, \mathfrak{R}_u$ | universal gas constant, $\mathrm{kg/(K \cdot mol) \cdot m^2/s^2}$ |
| $N_A$ | Avogadro's number, $\mathrm{1/mol}$ |
| $h$ | Planck's constant, $\mathrm{kg \cdot m^2/s}$ |
| $\hbar = h/(2\pi)$ | Planck's modified constant, $\mathrm{kg} \cdot \mathrm{m}^2/(\mathrm{s} \cdot \mathrm{rad})$ |
| $\mathcal{M}$ | Maxwellian velocity distribution |
| $\mathcal{E}$ | total energy distribution |
| $\boldsymbol{\mathcal{V}}$ | velocity vector distribution |
| $\mathcal{H}$ | Boltzmann's quantity (H-theorem) |
| $\mathbb{f}$ | general velocity distribution function |
| $\mathbb{f}^*$ | Hermite expanded distribution |
| $\mathbb{f}_\alpha^*, \mathbb{f}_\alpha^{\mathbb{m}}$ | discrete distribution on $\alpha$ quadrature node |
| $\psi_\alpha$ | Fourier wave form of $\mathcal{M}_\alpha^*$ |
| $\Psi_m$ | Fourier wave moment at the $m$th moment |
| $\mathbb{E}[\boldsymbol{\xi}^m] = \int_{-\infty}^{+\infty} \mathbb{f}\boldsymbol{\xi}^m \,\mathrm{d}\boldsymbol{\xi}$ | expectation or moment operation |
| $\Psi$ | quantum mechanical wave function |
| $\mu$ | 1st coefficient of viscosity, $\mathrm{kg}/(\mathrm{m} \cdot \mathrm{s})$ (also considered shear or dynamic viscosity) |
| $\zeta$ | 2nd coefficient of viscosity, $\mathrm{kg}/(\mathrm{m} \cdot \mathrm{s})$ |
| $\eta$ | total or bulk viscosity, $\mathrm{kg}/(\mathrm{m} \cdot \mathrm{s})$ (also considered volumetric or dilatational viscosity) |
| $\kappa$ | coefficient of thermal conductivity, $\mathrm{kg}/\mathrm{K} \cdot \mathrm{m}/\mathrm{s}^3$ |
| Pr | Prandtl number |
| $c_P$ | specific heat at constant pressure, $1/\mathrm{K} \cdot \mathrm{m}^2/\mathrm{s}^2$ |
| $c_V$ | specific heat at constant volume, $1/\mathrm{K} \cdot \mathrm{m}^2/\mathrm{s}^2$ |
| $a$ | constant related to bulk viscosity and adiabatic index |
| $\gamma$ | specific heat ratio or adiabatic index (monatomic gas) |
| $\Gamma$ | specific heat ratio or adiabatic index (polyatomic gas) |
| $b$ | constant related to thermal conductivity, $\mathrm{kg}/(\mathrm{m} \cdot \mathrm{s}^2)$ |
| $\sigma_{ij}$ | stress tensor, $\mathrm{kg}/(\mathrm{m} \cdot \mathrm{s}^2)$ |
| $q_i$ | heat flux vector, $\mathrm{kg}/\mathrm{s}^3$ |
| $\delta_{ij}, \mathbf{1}$ | Kronecker delta tensor[2] |
| $\epsilon_{ijk}$ | Levi-Civita or permutation tensor[3] |
| $\mathrm{e}\{x\}$ | exponential function |
| $\Xi_{ijkl}$ | tensor product: $\xi_i \xi_j \xi_k \xi_l$ |

## ABBREVIATIONS

| | |
|---|---|
| C-E | Chapman-Enskog |
| N-S | Navier-Stokes |
| N-S-F | Navier-Stokes-Fourier |
| ODE | ordinary differential equation |
| PDE | partial differential equation |
| VDF | velocity distribution function |
| CFD | computational fluid dynamics |
| DSMC | direct simulation Monte Carlo |
| LBM | lattice Boltzmann method |
| LBE | lattice Boltzmann equation |
| RV | random variable |
| FBE | freestream Boltzmann equation |
| FSE | freestream Schrödinger equation |

# THE MAXWELLIAN DISTRIBUTION

To describe the translational energy of a monatomic idealized gas, a velocity distribution function was first derived by Maxwell based on the assumption that each probability density for the three velocity components of an object are independent of each other [1] (Ch. 4.12). This assumption is sufficient for an inviscid flow (at the Euler level). To account for stress (at the N-S level), more measures need to be taken. Maxwell's "other equation" is referred to as the Maxwellian. Soon after, Boltzmann showed that by applying his H-theorem, Maxwell's VDF abides the second law of thermodynamics [1] (Ch. 4.1) meaning at uniform steady state, the quantity $\mathcal{H}$ will decreasingly approach a limit. A modernized derivation will be given as follows [2] (Ch. 7.3.4). Beginning with Boltzmann's entropy equation $s = -k \ln p$ [11] (Ch. 7.2) the probability function will be defined as $p = \mathrm{e}\{-\varepsilon/kT\}$. Inserting $p$ into $s$ gives the basic entropy definition $s \equiv \varepsilon/T$. The energy modeled is kinetic translational energy which can be described by the classical equation $\varepsilon = m(\xi - \mathrm{v})^2/2$, which is given in terms of a relative velocity. By feeding the energy into the probability function, a raw form of the Maxwellian is obtained. By assuming a continuous energy spectrum and normalizing, the one-dimensional Maxwellian can be derived which reveals a Gaussian with mean v and variance $\vartheta^2 = RT$. Below is the Maxwellian in $D$ Euclidean dimensions given in statistical form:

$$\mathcal{M}(\boldsymbol{\xi}; \mathbf{v}, \vartheta^2) = \prod_{i=1}^{D} \frac{1}{\sqrt{2\pi\vartheta^2}} \mathrm{e}\left\{-\frac{1}{2\vartheta^2}(\xi_i - \mathrm{v}_i)^2\right\} \tag{1}$$

The associated covariance matrix is noted as $\boldsymbol{\Lambda} = RT\mathbf{1}$. The distribution is then multiplied by $\rho$ which is the mass density of the gas or fluid per unit volume. A more commonly found form is:

$$\mathcal{M}(\boldsymbol{\xi}, \mathbf{v}, T, \rho) = \frac{\rho}{(2\pi RT)^{D/2}} \mathrm{e}\left\{-\frac{1}{2RT}(\boldsymbol{\xi} - \mathbf{v}) \cdot (\boldsymbol{\xi} - \mathbf{v})\right\} \tag{2}$$

Where $(\boldsymbol{\xi} - \mathbf{v}) \cdot (\boldsymbol{\xi} - \mathbf{v}) = \xi^2 - 2\xi_i \mathrm{v}_i + \mathrm{v}^2$. It will be noted that the above equation is a solution of the following partial differential equation which is of diffusion type:

$$\frac{\partial}{\partial T} \mathbb{f}(\boldsymbol{\xi}, T) = \frac{k}{2m} \frac{\partial^2}{\partial \xi_k \partial \xi_k} \mathbb{f}(\boldsymbol{\xi}, T) \tag{3}$$

Assuming homogenous boundary conditions allows for the separation of variables where the separation constant has units of per joule. Then assuming a continuous eigenvalue spectrum $\boldsymbol{\omega}$ over an infinite random velocity domain $\boldsymbol{\xi}$ gives the Fourier-type solution of the form:

$$\mathcal{M}(\boldsymbol{\xi}, \mathbf{v}, T, \rho) = \frac{\rho}{(2\pi)^D} \int_{-\infty}^{+\infty} \mathrm{e}\left\{-\frac{1}{2}(\boldsymbol{\omega} \cdot \boldsymbol{\omega})RT \pm i\boldsymbol{\omega} \cdot (\boldsymbol{\xi} - \mathbf{v})\right\} \mathrm{d}\boldsymbol{\omega} \tag{4}$$

See [6] (Ch. 10) for more details and $i$ is the imaginary number. An alternate derivation of the Maxwellian can be performed by using the maximum entropy principle. The objective is to minimize $\mathcal{H}$ using the Lagrange multiplier method in one dimension therefore assuming random velocity component independence:

$$\mathcal{H} = \min \int_{-\infty}^{+\infty} \mathbb{f}(\xi) \ln \mathbb{f}(\xi)\, \mathrm{d}\xi \tag{5}$$

Under the given constraints being the normalization condition, mean (velocity), and variance (specific thermal

energy) respectively:

$$1 = \int_{-\infty}^{+\infty} \mathbb{f}(\xi)\, \mathrm{d}\xi, \qquad \mathrm{v} = \int_{-\infty}^{+\infty} \mathbb{f}(\xi)\xi\, \mathrm{d}\xi, \qquad RT = \int_{-\infty}^{+\infty} \mathbb{f}(\xi)(\xi - \mathrm{v})^2\, \mathrm{d}\xi \tag{6}$$

The Lagrange function is formed as:

$$\mathcal{L} = \int_{-\infty}^{+\infty} \mathbb{f} \ln \mathbb{f}\, \mathrm{d}\xi + \mathrm{A} \int_{-\infty}^{+\infty} \mathbb{f}\, \mathrm{d}\xi + \mathrm{B} \int_{-\infty}^{+\infty} \mathbb{f}\xi\, \mathrm{d}\xi + \mathrm{C} \int_{-\infty}^{+\infty} \mathbb{f}(\xi - \mathrm{v})^2\, \mathrm{d}\xi \tag{7}$$

Setting $\partial\mathcal{L}/\partial\mathbb{f} = 0$ is the optimization condition giving:

$$\int_{-\infty}^{+\infty} (\ln \mathbb{f} + 1 + \mathrm{A} + \mathrm{B}\xi + \mathrm{C}(\xi - \mathrm{v})^2)\, \mathrm{d}\xi = 0 \tag{8}$$

Thus allowing for the isolation of $\mathbb{f}$ by setting the integrand equal to zero:

$$\mathbb{f} = \exp\{-\mathrm{B}\xi - \mathrm{C}(\xi - \mathrm{v})^2\} \tag{9}$$

Where terms that do not arise in the polynomial part of $\partial\mathbb{f}/\partial\xi$ are excluded. The constraints can be found in terms of the multipliers by generating moments:

$$1 = \exp\left\{\frac{\mathrm{B}^2}{4\mathrm{C}} - \mathrm{Bv}\right\}\sqrt{\frac{\pi}{\mathrm{C}}} \tag{10}$$

$$\mathrm{v} = \exp\left\{\frac{\mathrm{B}^2}{4\mathrm{C}} - \mathrm{Bv}\right\}\sqrt{\frac{\pi}{\mathrm{C}}}\left(\mathrm{v} - \frac{\mathrm{B}}{2\mathrm{C}}\right) \;\Rightarrow\; \mathrm{B} = 0 \tag{11}$$

$$RT = \exp\left\{\frac{\mathrm{B}^2}{4\mathrm{C}} - \mathrm{Bv}\right\}\sqrt{\frac{\pi}{\mathrm{C}}}\left(\frac{\mathrm{B}^2}{4\mathrm{C}^2} + \frac{1}{2\mathrm{C}}\right) \;\Rightarrow\; \mathrm{C} = \frac{1}{2RT} \tag{12}$$

Finding an expression for the velocity gives $\mathrm{B} = 0$. Then finding an expression for the specific thermal energy gives $\mathrm{C} = 1/(2RT)$. As a result, $\mathbb{f}$ becomes Maxwellian after normalizing which can be expanded to higher dimensions by use of the factorial in (1). The entropy can then be calculated as:

$$S = -nk \int_{-\infty}^{+\infty} \mathcal{M}(\boldsymbol{\xi}, \mathbf{v}, T) \ln \mathcal{M}(\boldsymbol{\xi}, \mathbf{v}, T, \rho)\, \mathrm{d}\boldsymbol{\xi} = nk \ln(2\pi RT)^{D/2} - nk \ln \rho + nkD/2 \tag{13}$$

The Gibbs relation excluding the change in internal energy and any chemical potentials gives $T\partial S = P\partial V$. Dividing both sides by $\partial V$ and utilizing the above entropy result gives the equation of state $PV = nkT$ where the mass density $\rho$ is $mn/V$ and the specific gas constant $R = k/m$. There are several ways to derive the Maxwellian, but as one can observe, it is simply a Gaussian or normal distribution. A unique feature of the Maxwellian (2) is the system level quantities obtained from the moments of the distribution taken over random velocity space. The moments bridge the gap between a theoretical probability space and measurements that can be taken in three-dimensional Euclidean space. The zeroth moment equates to the gas density due to the previously applied normalization condition. The first moment reveals the gas momentum density vector. The second scalar moment produces the scalar total energy density. To convert from temperature to pressure, the ideal gas law is used for the equation of state $P = \rho RT$. The moments are given by the general formula:

$$\mathbb{E}[\boldsymbol{\xi}^m] = \int_{-\infty}^{+\infty} \mathcal{M}(\boldsymbol{\xi})\boldsymbol{\xi}^m \, \mathrm{d}\boldsymbol{\xi}, \qquad m \in \mathbb{W} \tag{14}$$

Where the exponent is a tensor product operation. For example, the summational invariants are given as:

$$\mathbb{E}[\xi^0] = \int_{-\infty}^{+\infty} \mathcal{M}(\boldsymbol{\xi}) \, \mathrm{d}\boldsymbol{\xi}, \qquad \mathbb{E}[\xi_i] = \int_{-\infty}^{+\infty} \mathcal{M}(\boldsymbol{\xi})\xi_i \, \mathrm{d}\boldsymbol{\xi}, \qquad \mathbb{E}[\xi^2] = \int_{-\infty}^{+\infty} \mathcal{M}(\boldsymbol{\xi})\xi^2 \, \mathrm{d}\boldsymbol{\xi} \tag{15}$$

For deriving higher order moments, due to the increasing size of the tensors, it is beneficial to transform the integrals into derivatives using the moment generating function:

$$\mathbb{M}(\boldsymbol{\omega}) = \int_{-\infty}^{+\infty} \mathcal{M}(\boldsymbol{\xi}) \, \mathrm{e}\{\xi_i \omega_i\} \, \mathrm{d}\boldsymbol{\xi} \;\Rightarrow\; \mathbb{E}[\boldsymbol{\xi}^m] = \left.\frac{\partial^m \mathbb{M}}{\partial \boldsymbol{\omega}^m}\right|_{\boldsymbol{\omega}=0} \tag{16}$$

However, due to the properties of normal distributions, an even simpler operation was found to be:

$$\mathbb{E}[\boldsymbol{\xi}^m] = \frac{\rho(RT)^m}{f(\mathbf{v},T)} \frac{\partial^m}{\partial \mathbf{v}^m} f(\mathbf{v},T), \qquad f(\mathbf{v},T) = \frac{1}{(2\pi RT)^{D/2}} \mathrm{e}\left\{+\frac{1}{2RT}(\mathbf{v}\cdot\mathbf{v})\right\} \tag{17}$$

Where $f$ is defined as the pseudo-Maxwellian used strictly for the sake of deriving moments. Maple proved to be a viable software package for computing the integrals in integral form but the derivatives can be taken to higher order moments with more ease using a symbolic MATLAB algorithm. Below are the moments of the Maxwellian from zeroth to fourth order, see [3] (Ch. 2) for notation. From these moments, the scalar energy density and energy flux density vector can be found by taking partial moments. Beyond the primary moments are the central moments where the expectation values are taken with respect to the relative or peculiar velocity denoted $\boldsymbol{v} = \boldsymbol{\xi} - \mathbf{v}$. Odd order central moments equate to zero and the even orders do not contain advective quantities. The zeroth order central moment is equal to the zeroth order primary moment. The second order central moment results in the hydrostatic pressure. From the second order central moment, the variance can be found which is the sum of the hydrostatic pressures in all directions in $D$ Euclidean space and is referred to as the internal energy density. From the third order central moment, the heat flux density vector equates to zero meaning the Maxwellian is a velocity distribution function at thermodynamic equilibrium. The abbreviated specific thermal energy is denoted $\theta = RT$.

Total Primary Moments

$$\mathbb{E}[\xi^0] = \rho \;\Rightarrow\; \text{mass density} \tag{18}$$

$$\mathbb{E}[\xi_i] = \rho \mathrm{v}_i \;\Rightarrow\; \text{momentum density vector (mean)} \tag{19}$$

$$\mathbb{E}[\xi_i \xi_j] = \rho \mathrm{v}_i \mathrm{v}_j + P\delta_{ij} \;\Rightarrow\; \text{energy density tensor}^2 \tag{20}$$

$$\mathbb{E}[\xi_i \xi_j \xi_k] = \rho \mathrm{v}_i \mathrm{v}_j \mathrm{v}_k + P(\mathrm{v}_i \delta_{jk} + \mathrm{v}_j \delta_{ik} + \mathrm{v}_k \delta_{ij}) \;\Rightarrow\; \text{energy flux density tensor}^3 \tag{21}$$

$$\mathbb{E}\left[\xi_i\xi_j\xi_k\xi_l\right] = \rho \mathrm{v}_i\mathrm{v}_j\mathrm{v}_k\mathrm{v}_l + P\left(\mathrm{v}_i\mathrm{v}_j\delta_{kl} + \mathrm{v}_i\mathrm{v}_k\delta_{jl} + \mathrm{v}_i\mathrm{v}_l\delta_{jk} + \mathrm{v}_j\mathrm{v}_k\delta_{il} + \mathrm{v}_j\mathrm{v}_l\delta_{ik} + \mathrm{v}_k\mathrm{v}_l\delta_{ij}\right) + P\theta\delta_{ijkl} \quad (22)$$

<u>Partial Primary Moments</u>

$$\mathbb{E}[\xi^2] = \rho\mathrm{v}^2 + DP \;\Rightarrow\; \text{scalar energy density} \quad (23)$$

$$\mathbb{E}[\xi^2\xi_i] = (\rho\mathrm{v}^2 + (D+2)P)\mathrm{v}_i \;\Rightarrow\; \text{energy flux density vector} \quad (24)$$

$$\mathbb{E}\left[\xi^2\xi_i\xi_j\right] = (\rho\mathrm{v}^2 + (D+4)P)\mathrm{v}_i\mathrm{v}_j + P(\mathrm{v}^2 + (D+2)\theta)\delta_{ij} \quad (25)$$

<u>Partial Primary Moments</u> (rewritten using $\rho E = \rho\mathrm{v}^2 + DP \therefore E = \mathrm{v}^2 + D\theta$)

$$\mathbb{E}[\xi^2] = \rho E \quad (26)$$

$$\mathbb{E}[\xi^2\xi_i] = (\rho E + 2P)\mathrm{v}_i \quad (27)$$

$$\mathbb{E}\left[\xi^2\xi_i\xi_j\right] = (\rho E + 4P)\mathrm{v}_i\mathrm{v}_j + P(E + 2\theta)\delta_{ij} \quad (28)$$

<u>Total Central Moments</u>

$$\mathbb{E}[v_i] = 0 \quad (29)$$

$$\mathbb{E}\left[v_iv_j\right] = P\delta_{ij} \;\Rightarrow\; \text{hydrostatic pressure} \quad (30)$$

$$\mathbb{E}\left[v_iv_jv_k\right] = 0 \quad (31)$$

$$\mathbb{E}\left[v_iv_jv_kv_l\right] = P\theta\delta_{ijkl} \quad (32)$$

<u>Partial Central Moments</u>

$$\mathbb{E}[v^2] = DP \;\Rightarrow\; \text{internal energy density (variance)} \quad (33)$$

$$\mathbb{E}[v^2v_i] = 0 \;\Rightarrow\; \text{heat flux density vector} \quad (34)$$

$$\mathbb{E}\left[v^2v_iv_j\right] = (D+2)P\theta\delta_{ij} \quad (35)$$

<u>Mixed Moments</u>

$$\mathbb{E}\left[v_i\xi_j\right] = P\delta_{ij} \quad (36)$$

$$\mathbb{E}\left[\xi_i\xi_jv_k\right] = P\left(\mathrm{v}_i\delta_{jk} + \mathrm{v}_j\delta_{ik}\right) \quad (37)$$

$$\mathbb{E}\left[v_iv_j\xi_k\right] = P\mathrm{v}_k\delta_{ij} \quad (38)$$

$$\mathbb{E}\left[v_iv_j\xi_k\xi_l\right] = P\left(\mathrm{v}_k\mathrm{v}_l\delta_{ij} + \theta\delta_{ijkl}\right) \quad (39)$$

Below are Euler's system of partial differential equations which describe the transport of mass, linear momentum, and energy of an inviscid fluid where $\rho$, $\mathbf{v}$, and $P$ are considered dependent upon Eulerian coordinates (versus Lagrangian) in space and time:

Mass: $$\frac{\partial \rho}{\partial t} = -\frac{\partial \rho \mathrm{v}_i}{\partial x_i} \tag{40}$$

Momentum: $$\frac{\partial}{\partial t}\rho \mathrm{v}_i = -\frac{\partial}{\partial x_j}\left(\rho \mathrm{v}_i \mathrm{v}_j + P\delta_{ij}\right) \tag{41}$$

Energy: $$\frac{\partial}{\partial t}(\rho \mathrm{v}^2 + DP) = -\frac{\partial}{\partial x_i}(\rho \mathrm{v}^2 + (D+2)P)\mathrm{v}_i \tag{42}$$

$$\frac{\partial}{\partial t}(\rho E_{tr} + \rho E_{in}) = -\frac{\partial}{\partial x_i}(\rho E_{tr} + \rho H_{st})\mathrm{v}_i \tag{43}$$

In the energy equation, the specific translational energy is defined as $\mathrm{v}^2/2$, the specific internal energy is defined as $DRT/2$, and the specific static enthalpy as $E_{in} + RT$. The system (40-42) can be rearranged in terms of the material derivatives of the spacetime variables:

Density: $$\left(\frac{\partial}{\partial t} + \mathrm{v}_i \frac{\partial}{\partial x_i}\right)\rho = -\rho \frac{\partial \mathrm{v}_i}{\partial x_i} \tag{44}$$

Velocity: $$\left(\frac{\partial}{\partial t} + \mathrm{v}_k \frac{\partial}{\partial x_k}\right)\mathrm{v}_i = -\frac{1}{\rho}\frac{\partial P}{\partial x_i} \tag{45}$$

Pressure: $$\left(\frac{\partial}{\partial t} + \mathrm{v}_i \frac{\partial}{\partial x_i}\right)P = -\gamma P \frac{\partial \mathrm{v}_i}{\partial x_i} \tag{46}$$

Temperature: $$\left(\frac{\partial}{\partial t} + \mathrm{v}_i \frac{\partial}{\partial x_i}\right)T = -\frac{2T}{D}\frac{\partial \mathrm{v}_i}{\partial x_i} \tag{47}$$

Where $\gamma = (D+2)/D$ is considered the adiabatic index or heat capacity ratio as $\gamma = c_P/c_V$. One of the main focuses in this work is that the system (40-42) can also be rewritten in terms of the moments of the Maxwellian:

0th Moment: $$\frac{\partial}{\partial t}\mathbb{E}[\xi^0] = -\frac{\partial}{\partial x_i}\mathbb{E}[\xi_i] \tag{48}$$

1st Moment: $$\frac{\partial}{\partial t}\mathbb{E}[\xi_i] = -\frac{\partial}{\partial x_j}\mathbb{E}\left[\xi_i \xi_j\right] \tag{49}$$

2nd Scalar Moment: $$\frac{\partial}{\partial t}\mathbb{E}[\xi^2] = -\frac{\partial}{\partial x_i}\mathbb{E}[\xi^2 \xi_i] \tag{50}$$

This reformation of the Eulerian system is supporting evidence for the existence of the freestream Boltzmann equation which will be discussed further in a later section. Using Boltzmann's H-theorem, the material derivative of $\mathbb{f} \ln \mathbb{f}$ becomes zero at equilibrium [14]:

$$\frac{\partial}{\partial t}\int_{-\infty}^{+\infty} \mathcal{M} \ln \mathcal{M} \, \mathrm{d}\boldsymbol{\xi} + \frac{\partial}{\partial x_i}\int_{-\infty}^{+\infty} \mathcal{M}\xi_i \ln \mathcal{M} \, \mathrm{d}\boldsymbol{\xi} = 0 \tag{51}$$

Where Boltzmann's H-quantity is calculated as:

$$\mathcal{H} = \int_{-\infty}^{+\infty} \mathcal{M} \ln \mathcal{M} \, d\boldsymbol{\xi} = \rho \ln \rho - \rho \ln(2\pi RT)^{D/2} - \rho D/2 \tag{52}$$

Please note that both Maxwellians in $\mathcal{M} \ln \mathcal{M}$ are functions of $\rho$ unlike the equation of state derivation from the entropy equation (13). After expanding (51) and applying the continuity equation (40) one obtains:

$$\frac{1}{\rho}\left(\frac{\partial}{\partial t} + \mathrm{v}_i \frac{\partial}{\partial x_i}\right)\rho = \frac{D}{2T}\left(\frac{\partial}{\partial t} + \mathrm{v}_i \frac{\partial}{\partial x_i}\right)T \tag{53}$$

Substituting the material derivatives (44) and (47) into the above proves the equality relation in (51). The general H-theorem principle is that the following inequality should hold true:

$$\frac{\partial}{\partial t} \int_{-\infty}^{+\infty} \mathbb{f} \ln \mathbb{f} \, \mathrm{d}\boldsymbol{\xi} + \frac{\partial}{\partial x_i} \int_{-\infty}^{+\infty} \mathbb{f} \xi_i \ln \mathbb{f} \, \mathrm{d}\boldsymbol{\xi} \leq 0, \qquad \forall\, \mathbb{f} \geq 0 \tag{54}$$

The physics defined thus far can be considered classical mechanical. A general link can be formed to the more modern field of quantum mechanics by using mathematical methods taken from the derivation of the Schrödinger equation. To begin, this derivation will be briefly reviewed with the initial assumption being the existence of the freestream Boltzmann equation (77). The random velocity variable $\boldsymbol{\xi}$ will be converted into $\boldsymbol{\chi}/\tau$ where $\boldsymbol{\chi}$ and $\tau$ are some arbitrary finite change in position and time respectively. The freestream Boltzmann equation in $D$ Euclidean space is given as:

$$\frac{\partial}{\partial t}\Psi(\boldsymbol{x}, t) = -\frac{\chi_k}{\tau}\frac{\partial}{\partial x_k}\Psi(\boldsymbol{x}, t) \tag{55}$$

The variable $\Psi$ is the function aimed at being transformed into a velocity distribution function dependent upon space and time. It will be noted the above PDE is a distribution advection equation. Solutions are of exponential propagation type including both real and imaginary. Considering the imaginary type, the solution becomes of wave form due to Euler's formula:

$$\Psi(\boldsymbol{x}, t) = \mathbb{e}\left\{2\pi i\left(\frac{1}{D}\frac{x_k}{\chi_k} - \frac{t}{\tau}\right)\right\} \tag{56}$$

The de Broglie wave equation is $h = \boldsymbol{p} \cdot \boldsymbol{\chi}$ where the vector $\boldsymbol{p}$ is the momentum and $h$ is Planck's constant. It was recently mentioned that $\boldsymbol{\chi}$ is some change in position which can also be interpreted as a wave length associated with the momentum. Planck's law is $h = \varepsilon\tau$ where $\varepsilon$ is the scalar energy which is equal to Planck's constant multiplied by a temporal frequency. The momentum and energy will be held fixed in space and time. The de Broglie wave equation and Planck's law will now be fed into the above solution giving:

$$\Psi(\boldsymbol{x}, t) = \mathbb{e}\left\{\frac{i}{\hbar}\left(\frac{1}{D}p_k x_k - \varepsilon t\right)\right\} \tag{57}$$

Which is now in terms of some discrete momentum and energy and $\hbar$ is Planck's modified constant. Next, derivatives with respect to space and time will be taken to derive energy relations. The first derivative with respect to time and the second scalar spatial derivative or Laplacian are given respectively:

$$\varepsilon\Psi(\boldsymbol{x},t)=i\hbar\frac{\partial}{\partial t}\Psi(\boldsymbol{x},t),\qquad \frac{1}{D^2}\frac{p^2}{2m}\Psi(\boldsymbol{x},t)=-\frac{\hbar^2}{2m}\frac{\partial^2}{\partial x_k\partial x_k}\Psi(\boldsymbol{x},t) \tag{58}$$

The energy being modeled is translational kinetic energy which is given by the classical equation $\varepsilon=p^2/(2m)$. Feeding the above derivatives into the energy equation gives the freestream Schrödinger equation:

$$i\hbar\frac{\partial}{\partial t}\Psi(\boldsymbol{x},t)=-\frac{\hbar^2}{2m}\frac{\partial^2}{\partial x_k\partial x_k}\Psi(\boldsymbol{x},t) \tag{59}$$

Where $1/D^2$ has been dropped by reforming the FBE (55) as an average divergence directly proportional to the time rate of change. Now that this procedure has been established, an alternate route will be taken. The previous route will be considered quantum mechanical and the new route will be considered classical mechanical. In this next step, one can see where the two theories begin to split. Restarting with the freestream Boltzmann equation, a real solution will be given in the form:

$$\mathbb{f}(\boldsymbol{x},t)=\mathbb{e}\left\{\frac{1}{D}\frac{x_k}{\chi_k}-\frac{t}{\tau}\right\} \tag{60}$$

The above solution at first glance may not appear to vary significantly from (56), however the exclusion of the imaginary number $i$ makes this a propagation exponential solution which is no longer in wave form i.e., sine and cosine functions. In a similar fashion to deriving (57), the below result is obtained:

$$\mathbb{f}(\boldsymbol{x},t)=\mathbb{e}\left\{\frac{1}{h}\left(\frac{1}{D}p_i x_i-\varepsilon t\right)\right\} \tag{61}$$

To take an alternative path, momentum and energy relations will be formed. The momentum will be given in terms of the relative or peculiar velocity $\boldsymbol{v}=\boldsymbol{\xi}-\mathbf{v}$ therefore $\boldsymbol{p}=m\boldsymbol{v}$. The kinetic translational energy $\varepsilon$ then becomes $mv^2/2$. In statistical terms, Planck's constant becomes part of the variance as $h/m$ in the above distribution after normalizing over $\boldsymbol{\xi}$. Referring to the Maxwellian (1), the variance was found to be $kT/m$ where $k$ is Boltzmann's constant and $T$ is the discrete temperature. In this classical theory, the constant giving the action of a photon will be converted using the thermal energy relation $h=kT\tau$ thus giving the thermal action. This forms a relation between kinetic and thermal energy of a system at a point in space and time. Modifying (61) gives the kernel for the spacetime dependent velocity distribution:

$$\mathbb{f}(\boldsymbol{\xi},\boldsymbol{x},t)=\mathbb{e}\left\{-\frac{1}{2RT\tau}\left(v^2t-\frac{2}{D}v_i x_i\right)\right\} \tag{62}$$

To form a proper probability distribution function, the above result will be normalized over random velocity space $\boldsymbol{\xi}$ which produces a modified Maxwellian:

$$\widetilde{\mathcal{M}}(\boldsymbol{\xi},\boldsymbol{x},t)=\left(\frac{t}{2\pi RT\tau}\right)^{D/2}\mathbb{e}\left\{-\frac{1}{2RT\tau}\left(v^2t-\frac{2}{D}v_i x_i+\frac{1}{D^2}\frac{x^2}{t}\right)\right\} \tag{63}$$

Setting the position in the above result equal to zero at time $t=\tau=1$ gives exactly the Maxwellian (2). Please see the below figure for plots in both space and time. At any arbitrary point in space, the distribution approaches a Dirac delta function as one moves forward in time. There is a one-to-one linear correspondence between $x$ and $\xi$ at $t=1$. Beyond this point in time, the Gaussian begins to rotate and approaches a Dirac delta function at the mean velocity for all $x$. The plot to the right shows this "twisting" at $t=20$ s.

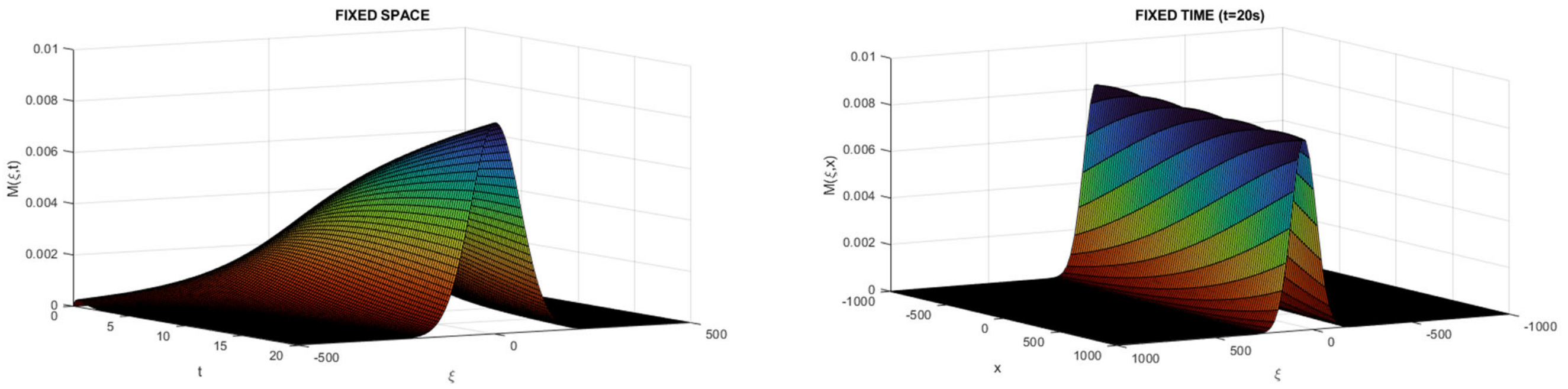

Figure #1 – Modified Maxwellian

Before continuing, for the sake of simplicity, a constant density flow will be considered where the incompressible assumption will be utilized. Therefore, the equation of state is lost so the expression $h/m = RT\tau$ will be converted into the kinematic viscosity $\nu$ which will be independent of time and space. Then $\nu$ will be set equal to $E\tau$ where $E$ will replace the specific thermal energy $RT$. For readability purposes, the relative velocity for this instance will be redefined as $\boldsymbol{c} = \boldsymbol{\xi} - \mathbf{v}$. The perturbation constant will be denoted $\epsilon = 1/D$. The above distribution (63) then becomes:

$$\widetilde{\mathcal{M}}(\boldsymbol{\xi}, \boldsymbol{x}, t) = \left(\frac{t}{2\pi\nu}\right)^{D/2} \mathrm{e}\left\{-\frac{1}{2\nu}\left(c^2 t - 2\epsilon c_i x_i + \frac{1}{t}\epsilon^2 x^2\right)\right\} \tag{64}$$

Generating the first and second moments of the modified distribution (64) while setting $t = \tau$ gives:

$$\widetilde{\mathbb{E}}[\xi_i] = \mathrm{v}_i \tag{65}$$

$$\widetilde{\mathbb{E}}[\xi_i \xi_j] = \mathrm{v}_i \mathrm{v}_j + E\delta_{ij} + \epsilon\hat{\xi}_i \mathrm{v}_j + \epsilon\hat{\xi}_j \mathrm{v}_i \tag{66}$$

The moments are truncated at $\mathcal{O}(\epsilon)$ and $\mathcal{O}(\epsilon^2)$ respectively. The independent stepping ratio is denoted $\hat{\boldsymbol{\xi}} = \boldsymbol{x}/t$. In a similar fashion to what was done earlier, the above moments will be substituted into the FBE. Before doing so, a velocity relation will be established by taking the first spatial derivative of (64) while setting the random velocity variable $\boldsymbol{\xi}$ equal to the mean velocity $\mathbf{v}$. This process gives the classical velocity operator:

$$\epsilon\hat{\xi}_i = \frac{1}{D}\frac{x_i}{t} = -D\nu\frac{\partial}{\partial x_i} \equiv -\nu\frac{\partial}{\partial x_i} \tag{67}$$

Where the constant $D$ is absorbed into the constant $\nu$. Feeding the moments (65) and (66) into the FBE (55) while feeding (67) into the second moment (66) gives the continuity $\partial \mathrm{v}_k/\partial x_k = 0$ and incompressible N-S momentum transfer equation:

$$\frac{\partial \mathrm{v}_i}{\partial t} + \mathrm{v}_k\frac{\partial \mathrm{v}_i}{\partial x_k} = \nu\frac{\partial^2 \mathrm{v}_i}{\partial x_k \partial x_k} \tag{68}$$

Insights into stress and diffusion are now obtained from this alternative treatment taken from mathematical methods used in the field of quantum mechanics. This new interpretation is an introduction to what will be done in the next section where the compressible N-S equations will be derived in a rigorous fashion. To end this section, Newton's equation of motion in one dimension will be derived in distribution form. The velocity evolution (68) in one dimension while applying the continuity equation and an external acceleration source is $d\mathrm{v}/dt = \mathrm{a}$. To begin the transformation, the total time derivative of an arbitrary distribution is found by applying the chain rule:

$$\frac{\mathrm{D}}{\mathrm{D}t}\mathbb{f}(\boldsymbol{\xi},\boldsymbol{x},t)=\frac{\partial \mathbb{f}}{\partial t}+\xi_i\frac{\partial \mathbb{f}}{\partial x_i}+\mathrm{a}_i\frac{\partial \mathbb{f}}{\partial \xi_i} \tag{69}$$

The governing partial differential equation becomes:

$$\frac{\partial}{\partial t}\mathcal{N}(\xi,t)=-\mathrm{a}\frac{\partial}{\partial \xi}\mathcal{N}(\xi,t) \tag{70}$$

For this particular case, the distribution will be defined as:

$$\mathcal{N}(\xi;\mathrm{v}(t),\sigma^2)=\frac{1}{(2\pi\sigma^2)^{D/2}}\mathrm{e}\left\{-\frac{1}{2\sigma^2}(\xi^2-2\xi\mathrm{v}+\mathrm{v}^2)\right\} \tag{71}$$

Where $\sigma^2$ is the variance of the normal distribution which will be become negligible for this instance. The velocity will be set dependent upon time as $\mathrm{v}(t)$. The derivative of (71) with respect to $\xi$ is calculated to be:

$$\sigma^2\frac{\partial}{\partial \xi}\mathcal{N}(\xi,t)=-v(t)\mathcal{N}(\xi,t) \tag{72}$$

Newton's second law in distribution form then becomes the following ordinary differential equation:

$$\sigma^2\dot{\mathcal{N}}=\mathrm{a}(\xi-\mathrm{v})\mathcal{N} \tag{73}$$

Setting the initial condition $\mathrm{v}(0)=U$ gives the solution $\mathrm{v}(t)=\mathrm{a}t+U$ which matches the original from solving $\dot{\mathrm{v}}=\mathrm{a}$ directly. In conclusion, an equivalent form of Newton's equation of motion in one dimension has been found in distribution form. The final distribution then becomes:

$$\mathcal{N}(\xi,t)=\frac{1}{(2\pi\sigma^2)^{D/2}}\mathrm{e}\left\{-\frac{1}{2\sigma^2}(\xi-U-\mathrm{a}t)^2\right\} \tag{74}$$

The above result is plotted below where to the right, it is shown that the velocity $\xi$ increases directly proportional to time $t$ through the applied acceleration constant a. To the left shows the temporal evolution of the distribution which follows a constant path throughout time.

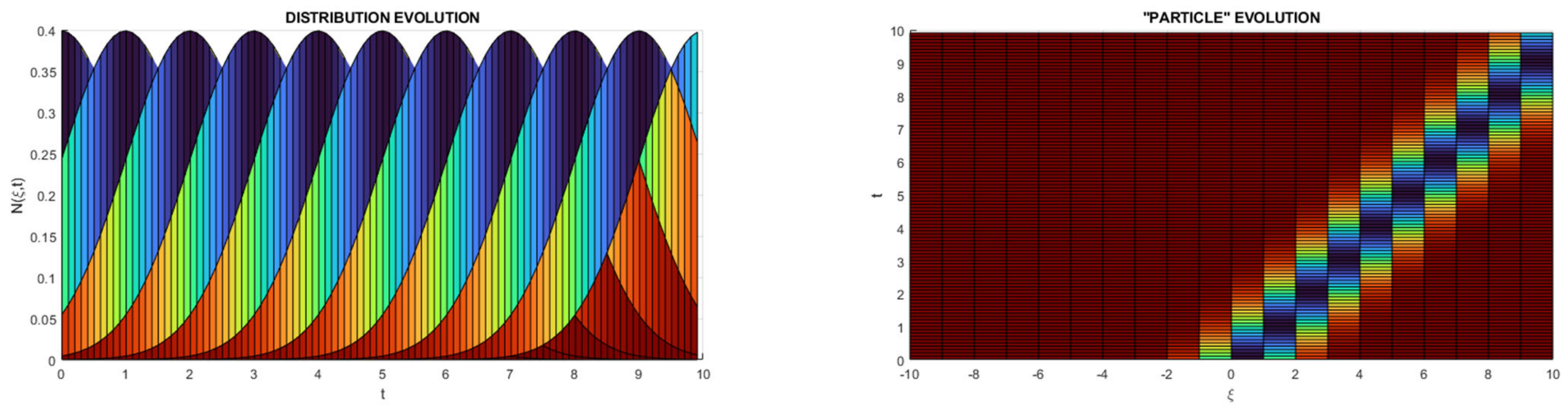

Figure #2 – Newton's 2nd Law

The difference between the first figure (from earlier) and the second (above) is that the first shows a distribution purely of space and time at constant velocity v. As time marches forwards, the probability that $\xi=\mathrm{v}$ increases because as mentioned earlier, the distribution approaches a Dirac delta function where the variance approaches zero. The second shows the affects on the distribution resulting from a changing velocity $\mathrm{v}(t)$ with respect to time. An interesting phenomena occurs when plotting. In order to plot, one has to discretize the domain of each variable. The above graph to the left was generated where $\Delta\xi > \Delta t$ where one can observe each discrete distribution located within each velocity cell at time $t$. This gives the illusion of "wave-like" behavior. However, a smooth result is obtained when $\Delta\xi \leq \Delta t$ where the discreteness of each distribution is not so apparent. Using a continuous or discrete set of velocity elements, the long-run behavior is unchanged. Veering back, to summarize, it has been introduced how system transport equations from the field of fluid mechanics can be derived from statistical mechanical principles. The next section will extend these concepts.

## TAYLOR SERIES AND TRANSPORT

The velocity distribution as a function of space and time will be denoted $\mathbb{f}(\boldsymbol{x},t)$. The following Eulerian and Navier-Stokes derivations are modified in reference to [7]. To begin, the VDF as some theoretical function of space and time will be expanded about some arbitrary point in time using a backward first order Taylor series:

$$-\frac{\mathbb{f}(\boldsymbol{x}-\boldsymbol{\xi}\Delta t, t-\Delta t)-\mathbb{f}(\boldsymbol{x},t)}{\Delta t} = \left(\frac{\partial}{\partial t}+\xi_i\frac{\partial}{\partial x_i}\right)\mathbb{f}(\boldsymbol{x},t) \tag{75}$$

The reason for using a backward difference will become clear in the second order expansion. Generating the zeroth, first, and second scalar moments of the above PDE while applying the laws of conservation of mass, momentum, and energy allows for the following assumption:

$$\int_{-\infty}^{+\infty}\mathbb{f}(\boldsymbol{x},t)\boldsymbol{\xi}^m\,\mathrm{d}\boldsymbol{\xi} = \int_{-\infty}^{+\infty}\mathbb{f}(\boldsymbol{x}-\boldsymbol{\xi}\Delta t, t-\Delta t)\boldsymbol{\xi}^m\,\mathrm{d}\boldsymbol{\xi}, \qquad m = 0:2 \tag{76}$$

The above relation holds true for the zeroth, first, and second scalar moments. Applying (76) to the moments of (75) at thermodynamic equilibrium, being $\mathbb{f} = \mathcal{M}$, gives the Eulerian transport system (40-42). This will be interpretated that the distribution function tied to the invariant moments is also conserved. At the zeroth moment, the freestream Boltzmann equation can be derived from this interpretation:

$$\frac{\partial}{\partial t}\mathbb{f}(\boldsymbol{x},t) = -\xi_i\frac{\partial}{\partial x_i}\mathbb{f}(\boldsymbol{x},t) \tag{77}$$

The law of conservation of distribution has thus been established. To begin to find a suitable evolution equation that will recover the N-S equations, the evolution in (75) will be expanded upon by taking a backward second order Taylor series expansion which gives:

$$-\frac{\mathbb{f}(\boldsymbol{x}-\boldsymbol{\xi}\Delta t, t-\Delta t)-\mathbb{f}(\boldsymbol{x},t)}{\Delta t} = \frac{\mathrm{D}}{\mathrm{D}t}\mathbb{f}(\boldsymbol{x},t) - \frac{\Delta t}{2}\frac{\mathrm{D}^2}{\mathrm{D}t^2}\mathbb{f}(\boldsymbol{x},t) \tag{78}$$

Where the operator $\mathrm{D}/\mathrm{D}t$ is the material derivative which is the right side of (75). In a similar fashion to the derivation of the freestream Boltzmann equation, through the use of applying the law of conservation of distribution, the above equation along with algebraic manipulation gives:

$$\left(\frac{\partial}{\partial t}+\xi_i\frac{\partial}{\partial x_i}\right)\mathbb{f}(\boldsymbol{x},t) = \frac{\Delta t}{2}\left(\frac{\partial^2}{\partial t^2}+2\xi_i\frac{\partial^2}{\partial x_i\partial t}+\xi_i\xi_j\frac{\partial^2}{\partial x_i\partial x_j}\right)\mathbb{f}(\boldsymbol{x},t) \tag{79}$$

Before continuing, it will be noted that fundamental PDE's arise from the freestream Boltzmann equation (77) which as it lies is an advection equation. Transforming $\boldsymbol{\xi}$ into an operator gives a diffusion equation as shown in

(67) and (236). Differentiating (77) with respect to time gives a relation for the mixed derivative in (79):

$$\frac{\partial^2}{\partial t^2}\mathbb{f}(\boldsymbol{x},t) = -\xi_i \frac{\partial^2}{\partial x_i \partial t}\mathbb{f}(\boldsymbol{x},t) \tag{80}$$

Substituting (77) and (80) into (79) allows for the derivation of the wave equation:

$$\frac{\partial^2}{\partial t^2}\mathbb{f}(\boldsymbol{x},t) = \xi_i\xi_j \frac{\partial^2}{\partial x_i \partial x_j}\mathbb{f}(\boldsymbol{x},t) \tag{81}$$

Restarting with (79) and substituting the above wave equation gives the N-S distribution transport PDE:

$$\left(\frac{\partial}{\partial t} + \xi_k \frac{\partial}{\partial x_k}\right)\mathbb{f}(\boldsymbol{x},t) = \Delta t \xi_j \frac{\partial}{\partial x_j}\left(\frac{\partial}{\partial t} + \xi_k \frac{\partial}{\partial x_k}\right)\mathbb{f}(\boldsymbol{x},t) \tag{82}$$

The right side of (82) will be shown to produce the stress tensor in the momentum equation and heat flux vector in the energy equation. As can be seen, the spatial derivative on the right operates on the material derivative of the $m + 1$ moment versus the $m^{\text{th}}$ moment on the left. At the zeroth moment, the right side equates to zero due to the relation from the Euler level momentum equation (41), thus reducing the entire PDE down to the mass transport or continuity equation (40). At the first moment is:

$$\frac{\partial}{\partial t}\mathbb{E}[\xi_i] + \frac{\partial}{\partial x_j}\mathbb{E}[\xi_i\xi_j] = \Delta t \frac{\partial}{\partial x_j}\left(\frac{\partial}{\partial t}\mathbb{E}[\xi_i\xi_j] + \frac{\partial}{\partial x_k}\mathbb{E}[\xi_i\xi_j\xi_k]\right) \tag{83}$$

The difference here is that the right side of (83) is not fully conserved as shown from the Euler relations (40-42). Upon substituting the second and third order primary moments (20) and (21) into the right side of (83), the stress tensor in raw form is formed as:

$$\sigma_{ij} = \Delta t \left[\frac{\partial}{\partial t}\left(\rho \mathrm{v}_i \mathrm{v}_j + P\delta_{ij}\right) + \frac{\partial}{\partial x_k}\left(\rho \mathrm{v}_i \mathrm{v}_j \mathrm{v}_k + P\mathrm{v}_i\delta_{jk} + P\mathrm{v}_j\delta_{ik} + P\mathrm{v}_k\delta_{ij}\right)\right] \tag{84}$$

The objective is to now resolve the temporal derivative in (84) into spatial derivatives. To do this, simple modifications to the density and velocity material derivatives, (44) and (45), are utilized giving:

$$\frac{\partial \rho}{\partial t} = -\mathrm{v}_k \frac{\partial \rho}{\partial x_k} - \rho \frac{\partial \mathrm{v}_k}{\partial x_k}, \qquad \frac{\partial \mathrm{v}_i}{\partial t} = -\mathrm{v}_k \frac{\partial \mathrm{v}_i}{\partial x_k} - \frac{1}{\rho}\frac{\partial P}{\partial x_i} \tag{85}$$

The temporal derivative of $\rho\mathbf{v}\otimes\mathbf{v}$ is then reformed into spatial derivatives:

$$\frac{\partial}{\partial t}\rho \mathrm{v}_i \mathrm{v}_j = -\frac{\partial}{\partial x_k}\left(\rho \mathrm{v}_i \mathrm{v}_j \mathrm{v}_k\right) - \mathrm{v}_i \frac{\partial P}{\partial x_j} - \mathrm{v}_j \frac{\partial P}{\partial x_i} \tag{86}$$

To transform the pressure time derivative, the pressure material derivative (46) will be used:

$$\frac{\partial P}{\partial t} = -\mathrm{v}_k \frac{\partial P}{\partial x_k} - \gamma P \frac{\partial \mathrm{v}_k}{\partial x_k} \tag{87}$$

Substituting (86) and (87) into (84) results in the following momentum equation:

$$\frac{\partial}{\partial t}\rho \mathrm{v}_i + \frac{\partial}{\partial x_j}\left(\rho \mathrm{v}_i \mathrm{v}_j + P\delta_{ij}\right) = \frac{\partial}{\partial x_j} P\Delta t \left(\frac{\partial \mathrm{v}_i}{\partial x_j} + \frac{\partial \mathrm{v}_j}{\partial x_i} - \frac{2}{D}\frac{\partial \mathrm{v}_k}{\partial x_k}\delta_{ij}\right) \tag{88}$$

Which is the Navier-Stokes momentum equation of a monatomic ideal gas where the bulk viscosity is zero thus falling under Stokes' hypothesis. The first coefficient of viscosity or shear viscosity, second coefficient of viscosity, and total or bulk viscosity is given respectively:

1st coefficient of viscosity: $\mu = P\Delta t$ (89)

2nd coefficient of viscosity: $\zeta = -2\mu/3$ (90)

Bulk viscosity: $\eta = 0$ (91)

The second coefficient is found from the relation $\eta = 2\mu/D + \zeta$ [10] (Ch. 45.4). The stress tensor is then:

$$\sigma_{ij} = \mu\left(\frac{\partial \mathrm{v}_i}{\partial x_j} + \frac{\partial \mathrm{v}_j}{\partial x_i} - \frac{2}{D}\frac{\partial \mathrm{v}_k}{\partial x_k}\delta_{ij}\right) \tag{92}$$

To recover a bulk viscosity, more measures need to be taken which will be covered in the next section. The next task will be to derive the energy transport equation. Taking the second scalar moment of (82) gives:

$$\frac{\partial}{\partial t}\mathbb{E}[\xi^2] + \frac{\partial}{\partial x_i}\mathbb{E}[\xi^2\xi_i] = \Delta t\frac{\partial}{\partial x_i}\left(\frac{\partial}{\partial t}\mathbb{E}[\xi^2\xi_i] + \frac{\partial}{\partial x_j}\mathbb{E}\left[\xi^2\xi_i\xi_j\right]\right) \tag{93}$$

Where the right side yields:

$$\Delta t\frac{\partial}{\partial x_i}\left[\frac{\partial}{\partial t}(\rho \mathrm{v}^2\mathrm{v}_i + (D+2)P\mathrm{v}_i) + \frac{\partial}{\partial x_j}\left(\rho \mathrm{v}^2\mathrm{v}_i\mathrm{v}_j + P\mathrm{v}^2\delta_{ij} + (D+4)P\mathrm{v}_i\mathrm{v}_j + (D+2)P\theta\delta_{ij}\right)\right] \tag{94}$$

In a similar fashion to the momentum derivation, the time derivative of $\rho \mathrm{v}^2\mathbf{v}$ becomes:

$$\frac{\partial}{\partial t}\rho \mathrm{v}^2\mathrm{v}_i = -\frac{\partial}{\partial x_j}\rho \mathrm{v}^2\mathrm{v}_i\mathrm{v}_j - 2\mathrm{v}_i\mathrm{v}_j\frac{\partial P}{\partial x_j} - \mathrm{v}^2\frac{\partial P}{\partial x_i} \tag{95}$$

The time derivative of $P\mathbf{v}$ is transformed into:

$$\frac{\partial}{\partial t}P\mathrm{v}_i = -\frac{2P\mathrm{v}_i}{D}\frac{\partial \mathrm{v}_k}{\partial x_k} - \theta\frac{\partial P}{\partial x_i} - \frac{\partial}{\partial x_j}P\mathrm{v}_i\mathrm{v}_j \tag{96}$$

Combining the last two equations and substituting into (94) produces the energy equation:

$$\frac{1}{2}\frac{\partial}{\partial t}(\rho \mathrm{v}^2 + DP) + \frac{1}{2}\frac{\partial}{\partial x_i}(\rho \mathrm{v}^2 + \gamma DP)\mathrm{v}_i = \frac{\partial}{\partial x_j}\mu\left[\left(\frac{\partial \mathrm{v}_i}{\partial x_j} + \frac{\partial \mathrm{v}_j}{\partial x_i}\right)\mathrm{v}_i - \frac{2}{D}\frac{\partial \mathrm{v}_k}{\partial x_k}\mathrm{v}_j\right] + \frac{\partial}{\partial x_j}\frac{\mu\gamma D}{2}\frac{\partial \theta}{\partial x_j} \tag{97}$$

Where the coefficient of thermal conductivity is recognized to be $\kappa = \mu\gamma DR/2$. The heat flux vector, translational energy density, internal energy density, and static enthalpy density are defined respectively:

Heat flux vector: $q_j = \kappa\partial T/\partial x_j$ (98)

Kinetic energy density: $\rho E_{tr} = \rho \mathrm{v}^2/2$ (99)

Internal energy density: $\rho E_{in} = DP/2$ (100)

Static enthalpy density: $\rho H_{st} = \rho E_{in} + P = \gamma DP/2 = (D+2)P/2$ (101)

As a result, the energy transport equation can be rewritten in the following form:

$$\frac{\partial}{\partial t}(\rho E_{tr} + \rho E_{in}) + \frac{\partial}{\partial x_i}(\rho E_{tr} + \rho H_{st})\mathrm{v}_i = \frac{\partial}{\partial x_j}\left(\sigma_{ij}\mathrm{v}_i + q_j\right) \tag{102}$$

Thermodynamically, the ideal gas equation of state was used from the beginning therefore, for a thermally perfect gas, the specific heat at constant pressure and volume are given respectively:

Specific heat at constant $P$: $c_P = \partial H_{st}/\partial T = \gamma DR/2 = (D+2)R/2$ (103)

Specific heat at constant $V$: $c_V = \partial E_{in}/\partial T = DR/2$ (104)

As a result, the Prandtl number is calculated to be $\mathrm{Pr} = c_P\mu/\kappa = 1$ because the coefficient of thermal conductivity is $\kappa = \mu c_P$. To recover a Prandtl number where the rate of momentum diffusion varies from the rate of thermal diffusion, more measures will need to be taken. This will be accomplished in the next section.

## TOTAL ENERGY DISTRIBUTION

Transport equations for a monatomic ideal gas have been derived from a translational energy distribution function where in this section the extension to polyatomic gases will be made. To include a bulk viscosity and adjustable Prandtl number and specific heat ratio, a separate total energy distribution function will be utilized which is a modified version of the one provided in [12]:

$$\mathcal{E}(\boldsymbol{\xi},\mathbf{v},T,\rho) = \frac{\rho\xi^2}{(2\pi RT)^{D/2}}\mathrm{e}\left\{-\frac{1}{2RT}v^2\right\} + \frac{aP}{(2\pi RT)^{D/2}}\mathrm{e}\left\{-\frac{1}{2RT}v^2\right\} + \frac{b}{(2\pi RT)^{D/2}}\mathrm{e}\left\{-\frac{1}{2RT}\xi^2\right\} \tag{105}$$

Or more simply:

$$\mathcal{E} = \rho\xi^2\mathcal{M}(\boldsymbol{\xi},\mathbf{v},T) + aP\mathcal{M}(\boldsymbol{\xi},\mathbf{v},T) + b\mathcal{M}(\boldsymbol{\xi},T) \tag{106}$$

The constant $a$ has been introduced to provide a bulk viscosity and adjustable specific heat ratio. The constant $b$ has been introduced to provide an adjustable Prandtl number. The total energy distribution above is a finite perturbation series expansion of the Maxwellian and allows for the inclusion of other modes of internal energy. The moments of (106) can be expressed as:

$$\int_{-\infty}^{+\infty} \mathcal{E}\boldsymbol{\xi}^m \,\mathrm{d}\boldsymbol{\xi} = \rho\,\mathrm{tr}\,\mathbb{E}[\boldsymbol{\xi}^{m+2}] + aP\mathbb{E}[\boldsymbol{\xi}^m] + b\mathbb{E}[\boldsymbol{v}^m] \tag{107}$$

The zeroth, first, and second order primary moments are given as follows:

$$\mathbb{E}_{\mathbb{1}}[\xi^0] = \rho\mathrm{v}^2 + DP + aP + b \tag{108}$$

$$\mathbb{E}_{\mathbb{1}}[\xi_i] = \rho \mathrm{v}^2 \mathrm{v}_i + (D+2)P\mathrm{v}_i + aP\mathrm{v}_i \tag{109}$$

$$\mathbb{E}_{\mathbb{1}}[\xi_i\xi_j] = (\rho \mathrm{v}^2 + (D+4)P)\mathrm{v}_i\mathrm{v}_j + P(\mathrm{v}^2 + (D+2)\theta)\delta_{ij} + aP(\mathrm{v}_i\mathrm{v}_j + \theta\delta_{ij}) + b\theta\delta_{ij} \tag{110}$$

At the Euler level, the transport PDE's are given as:

$$\frac{\partial}{\partial t}\mathbb{E}_{\mathbb{0}}[\xi^0] = -\frac{\partial}{\partial x_i}\mathbb{E}_{\mathbb{0}}[\xi_i] \;\Rightarrow\; \frac{\partial \rho}{\partial t} = -\frac{\partial \rho \mathrm{v}_i}{\partial x_i} \tag{111}$$

$$\frac{\partial}{\partial t}\mathbb{E}_{\mathbb{0}}[\xi_i] = -\frac{\partial}{\partial x_j}\mathbb{E}_{\mathbb{0}}[\xi_i\xi_j] \;\Rightarrow\; \frac{\partial}{\partial t}\rho \mathrm{v}_i = -\frac{\partial}{\partial x_j}(\rho \mathrm{v}_i\mathrm{v}_j + P\delta_{ij}) \tag{112}$$

$$\frac{\partial}{\partial t}\mathbb{E}_{\mathbb{1}}[\xi^0] = -\frac{\partial}{\partial x_i}\mathbb{E}_{\mathbb{1}}[\xi_i] \;\Rightarrow\; \frac{\partial}{\partial t}(\rho \mathrm{v}^2 + (D+a)P) = -\frac{\partial}{\partial x_i}(\rho \mathrm{v}^2 + (D+2+a)P)\mathrm{v}_i \tag{113}$$

The expectations $\mathbb{E}_{\mathbb{0}}$ refer to the moments of the Maxwellian and the expectations $\mathbb{E}_{\mathbb{1}}$ refer to the moments of the total energy distribution. The density and velocity time derivatives (85) remain the same however the pressure evolution becomes:

$$\frac{\partial P}{\partial t} = -\mathrm{v}_k\frac{\partial P}{\partial x_k} - \Gamma P\frac{\partial \mathrm{v}_k}{\partial x_k} \tag{114}$$

The constant $\Gamma$ is defined as $(A+2)/A$ where $A = D + a$. Going back to (84) and substituting the newly derived pressure material derivative gives the following momentum transport equation:

$$\frac{\partial}{\partial t}\rho \mathrm{v}_i + \frac{\partial}{\partial x_j}(\rho \mathrm{v}_i\mathrm{v}_j + P\delta_{ij}) = \frac{\partial}{\partial x_j}P\Delta t\left(\frac{\partial \mathrm{v}_i}{\partial x_j} + \frac{\partial \mathrm{v}_j}{\partial x_i} - \frac{2}{D}\frac{\partial \mathrm{v}_k}{\partial x_k}\delta_{ij} + \frac{2a}{DA}\frac{\partial \mathrm{v}_k}{\partial x_k}\delta_{ij}\right) \tag{115}$$

Once again, the first coefficient of viscosity or shear viscosity is recognized as $\mu = P\Delta t$ however the bulk viscosity is no longer zero. The new coefficients of viscosity are given below:

1st coefficient of viscosity: $\mu = P\Delta t$ (116)

2nd coefficient of viscosity: $\zeta = -2\mu/A$ (117)

Bulk viscosity: $\eta = 2\mu a/(DA)$ (118)

The final form of the N-S momentum transport equation then becomes:

$$\frac{\partial}{\partial t}\rho \mathrm{v}_i + \frac{\partial}{\partial x_j}(\rho \mathrm{v}_i\mathrm{v}_j + P\delta_{ij}) = \frac{\partial}{\partial x_j}\left[\mu\left(\frac{\partial \mathrm{v}_i}{\partial x_j} + \frac{\partial \mathrm{v}_j}{\partial x_i} - \frac{2}{D}\frac{\partial \mathrm{v}_k}{\partial x_k}\delta_{ij}\right) + \eta\frac{\partial \mathrm{v}_k}{\partial x_k}\delta_{ij}\right] \tag{119}$$

Where the new stress tensor is identified as:

$$\sigma_{ij} = \mu\left(\frac{\partial \mathrm{v}_i}{\partial x_j} + \frac{\partial \mathrm{v}_j}{\partial x_i} - \frac{2}{D}\frac{\partial \mathrm{v}_k}{\partial x_k}\delta_{ij}\right) + \eta\frac{\partial \mathrm{v}_k}{\partial x_k}\delta_{ij} \tag{120}$$

As a result, the complete N-S momentum transport equation has been recovered. To note, at the Euler level, the mass and momentum equations utilize the Maxwellian. However, the stress tensor in the momentum equation utilizes the pressure time derivative (114) which comes from the Eulerian energy equation which takes moments of the total energy distribution. The N-S energy transport will be modeled utilizing the total energy distribution:

$$\frac{\partial}{\partial t}\mathbb{E}_{\mathbb{1}}[\xi^0] + \frac{\partial}{\partial x_i}\mathbb{E}_{\mathbb{1}}[\xi_i] = \Delta t\frac{\partial}{\partial x_i}\left(\frac{\partial}{\partial t}\mathbb{E}_{\mathbb{1}}[\xi_i] + \frac{\partial}{\partial x_j}\mathbb{E}_{\mathbb{1}}[\xi_i\xi_j]\right) \tag{121}$$

The total scalar energy density is defined as $\rho E = \rho \mathrm{v}^2 + AP$ so for convenience, the previously calculated moments (108-110) will be redefined in terms of the total energy:

$$\mathbb{E}_{\mathbb{1}}[\xi^0] = \rho E + b \tag{122}$$

$$\mathbb{E}_{\mathbb{1}}[\xi_i] = (\rho E + 2P)\mathrm{v}_i \tag{123}$$

$$\mathbb{E}_{\mathbb{1}}[\xi_i\xi_j] = (\rho E + 4P)\mathrm{v}_i\mathrm{v}_j + P(E + 2\theta)\delta_{ij} + b\theta\delta_{ij} \tag{124}$$

Inserting the above moments into the N-S energy transport equation (121) yields on the right side:

$$\Delta t\frac{\partial}{\partial x_i}\left[\frac{\partial}{\partial t}(\rho E\mathrm{v}_i + 2P\mathrm{v}_i) + \frac{\partial}{\partial x_j}\left(\rho E\mathrm{v}_i\mathrm{v}_j + 4P\mathrm{v}_i\mathrm{v}_j + PE\delta_{ij} + 2P\theta\delta_{ij} + b\theta\delta_{ij}\right)\right] \tag{125}$$

Reforming Euler's total energy transport equation (113):

$$\frac{\partial}{\partial t}\rho E = -\frac{\partial}{\partial x_i}(\rho E\mathrm{v}_i + 2P\mathrm{v}_i) \tag{126}$$

Allowing for the temporal to spatial derivative transformations:

$$\frac{\partial}{\partial t}\rho E\mathrm{v}_i = -\frac{\partial}{\partial x_k}\rho E\mathrm{v}_i\mathrm{v}_k - E\frac{\partial P}{\partial x_i} - 2\mathrm{v}_i\frac{\partial}{\partial x_k}P\mathrm{v}_k \tag{127}$$

$$\frac{\partial}{\partial t}P\mathrm{v}_i = -\mathrm{v}_k\frac{\partial}{\partial x_k}P\mathrm{v}_i - \Gamma P\mathrm{v}_i\frac{\partial \mathrm{v}_k}{\partial x_k} - \theta\frac{\partial P}{\partial x_i} \tag{128}$$

Substituting the above temporal derivatives into (125) and dividing by two gives:

$$\frac{\partial}{\partial x_i}P\Delta t\left[\frac{\partial \mathrm{v}_i}{\partial x_k}\mathrm{v}_k + \frac{\partial \mathrm{v}_k}{\partial x_i}\mathrm{v}_k - \frac{2}{D}\frac{\partial \mathrm{v}_k}{\partial x_k}\mathrm{v}_i + \frac{2a}{DA}\frac{\partial \mathrm{v}_k}{\partial x_k}\mathrm{v}_i + \frac{AP + 2P + b}{2P}\frac{\partial \theta}{\partial x_i}\right] \tag{129}$$

The new coefficient of thermal conductivity is recognized as $\kappa = (\Gamma A\mu + b\Delta t)R/2$. The final energy becomes:

$$\frac{1}{2}\frac{\partial}{\partial t}\rho E + \frac{1}{2}\frac{\partial}{\partial x_i}(\rho E + 2P)\mathrm{v}_i = \frac{\partial}{\partial x_i}\left[\mu\left(\frac{\partial \mathrm{v}_i}{\partial x_k}\mathrm{v}_k + \frac{\partial \mathrm{v}_k}{\partial x_i}\mathrm{v}_k - \frac{2}{D}\frac{\partial \mathrm{v}_k}{\partial x_k}\mathrm{v}_i\right) + \eta\frac{\partial \mathrm{v}_k}{\partial x_k}\mathrm{v}_i + \kappa\frac{\partial T}{\partial x_i}\right] \tag{130}$$

Thus, recovering the complete N-S energy transport equation. The heat flux vector can be recognized in terms of the static enthalpy density and the constant $b$:

$$q_i = \kappa\frac{\partial T}{\partial x_i} = R\Delta t\left(\rho H_{st} + \frac{1}{2}b\right)\frac{\partial T}{\partial x_i} \tag{131}$$

Where the internal energy density and static enthalpy density are defined respectively:

$$\rho E_{in} = AP/2 = (D + a)P/2 \tag{132}$$

$$\rho H_{st} = \Gamma AP/2 = (D + a + 2)P/2 \tag{133}$$

The new specific heat at constant pressure and volume are calculated respectively:

$$c_P = \partial H_{st}/\partial T = \Gamma AR/2 = (D + a + 2)R/2 \tag{134}$$

$$c_V = \partial E_{in}/\partial T = AR/2 = (D + a)R/2 \tag{135}$$

The coefficient of thermal conductivity $\kappa = \mu c_P + bR\Delta t/2$ produces the resulting Prandtl number:

$$\Pr = \frac{\Gamma AP}{\Gamma AP + b} = \frac{(D + a + 2)P}{(D + a + 2)P + b} \tag{136}$$

Which is defined by the constants $a$ and $b$ and the pressure function $P$. The constant $a$ is dimensionless, related to the internal modes of energy, and contributes to the coefficient of bulk viscosity. The constant $b$ has units of pressure and contributes to the coefficient of thermal conductivity. To summarize, a more general form of the Navier-Stokes equations have been derived from statistical mechanical principles through the use of a coupled double distribution function approach. The main conclusion is that (119) and (130) can be derived from coupled distribution evolution equations through a multiscale moment analysis of (77) and (82). It was shown that this method allows for the extension to polyatomic gases which includes a bulk viscosity coefficient $\eta$ in the momentum transport, allows for an adjustable specific heat ratio $\Gamma$, and accounts for different rates of momentum versus thermal diffusion due to the calculated Prandtl number in (136). In addition, typically $\mu$, $\eta$, and $\kappa$ are empirically determined constants however their spacetime dependence has been shown from the right side of (115) and (129). Depending on the material, the constants $a$ and $b$ will need to be empirically found. It is clear how to determine $a$ based on the adiabatic index however more research needs to be done to uncover values for $b$ which appears would come from some theoretical heat reservoir. Lastly, all the derived coefficients are dependent upon the dissipation time $\Delta t$ which will be derived in future work.

## HERMITE SERIES AND VELOCITY DISCRETIZATION

Steering back towards the traditional Maxwellian (2), the objective is to now transform the continuous integrals in (48-50) into finite summations and the continuous random variable $\boldsymbol{\xi}$ into a discrete set. To accomplish this, the Maxwellian will be expanded using the Hermite series expansion, see [4] (Ch. 3). The below series transforms the exponential function $\mathcal{M}$ into a polynomial function $\mathcal{M}^*$ in terms of its moments:

$$\mathcal{M}^*(\boldsymbol{\xi}, \mathbf{v}, T, \rho) = \omega(\boldsymbol{\xi}) \sum_{n=0}^{N} \frac{1}{n!} \boldsymbol{A}^{(n)}(\mathbf{v}, T, \rho) \cdot \boldsymbol{H}^{(n)}(\boldsymbol{\xi}) \tag{137}$$

Where $\mathcal{M}^*$ is the truncated Maxwellian, $\omega$ is the weight function, $\boldsymbol{A}^{(n)}$ is the set of Hermite coefficients or moments, and $\boldsymbol{H}^{(n)}$ is the set of Hermite polynomials. The operation $\boldsymbol{A}^{(n)} \cdot \boldsymbol{H}^{(n)}$ is a full contraction. The Hermite series expansion will be used alongside the Gauss-Hermite quadrature transformation which transforms the integral of the product between the weight function $\omega$ and a polynomial function into a finite sum where the independent variable and weight function can be discretized into a discrete set:

$$\int_{-\infty}^{+\infty} \omega(\boldsymbol{\xi})\mathcal{P}(\boldsymbol{\xi})\,\mathrm{d}\boldsymbol{\xi} = \sum_{\alpha=1}^{Q} w_\alpha(c)\mathcal{P}(c\boldsymbol{e}_\alpha) \tag{138}$$

Where $\boldsymbol{\xi}_\alpha = c\boldsymbol{e}_\alpha$ is established as a discrete set of random variable velocities or lattices, $c$ is the lattice constant, the discrete set of weights is denoted $w_\alpha$ which is a function of the lattice constant, and $\mathcal{P}$ is a polynomial function. The Hermite polynomials and weight function are defined as:

$$\boldsymbol{H}^{(n)}(\boldsymbol{\xi}) = (-1)^n \frac{1}{\omega(\boldsymbol{\xi})}\frac{\partial^n}{\partial\boldsymbol{\xi}^n}\omega(\boldsymbol{\xi}), \qquad \omega(\boldsymbol{\xi}) = \prod_{i=1}^{D}\frac{1}{\sqrt{2\pi}}\mathrm{e}\left\{-\frac{1}{2}\xi_i^2\right\} \tag{139}$$

Where each order $n$ defines the dimensionality of the Hermite polynomials. The operation $\partial^n/\partial\boldsymbol{\xi}^n$ is a tensor product. So, at $n = 0$ the Hermite polynomial is a scalar, at $n = 1$ a vector, $n = 2$ a tensor[2], and so on. The number of equations per order $n$ is $D^n$. The weight function is a Gaussian with mean 0 and variance 1. The first five Hermite polynomials are given as:

$$H^{(0)} = 1 \tag{140}$$

$$H_i^{(1)} = \xi_i \tag{141}$$

$$H_{ij}^{(2)} = \xi_i\xi_j - \delta_{ij} \tag{142}$$

$$H_{ijk}^{(3)} = \xi_i\xi_j\xi_k - \left(\xi_i\delta_{jk} + \xi_j\delta_{ik} + \xi_k\delta_{ij}\right) \tag{143}$$

$$H_{ijkl}^{(4)} = \xi_i\xi_j\xi_k\xi_l - \left(\xi_i\xi_j\delta_{kl} + \xi_i\xi_k\delta_{jl} + \xi_i\xi_l\delta_{jk} + \xi_j\xi_k\delta_{il} + \xi_j\xi_l\delta_{ik} + \xi_k\xi_l\delta_{ij}\right) + \delta_{ijkl} \tag{144}$$

The Hermite coefficients or moments become independent of $\boldsymbol{\xi}$ and defined as:

$$\boldsymbol{A}^{(n)}(\mathbf{v}, T, \rho) = \int_{-\infty}^{+\infty} \mathcal{M}(\boldsymbol{\xi}, \mathbf{v}, T, \rho)\boldsymbol{H}^{(n)}(\boldsymbol{\xi})\,\mathrm{d}\boldsymbol{\xi} \tag{145}$$

The Hermite coefficients resemble the moments of the Maxwellian:

$$A^{(0)} = \mathbb{E}^{(0)} \tag{146}$$

$$A_i^{(1)} = \mathbb{E}_i^{(1)} \tag{147}$$

$$A_{ij}^{(2)} = \mathbb{E}_{ij}^{(2)} - \mathbb{E}^{(0)}\delta_{ij} \tag{148}$$

$$A_{ijk}^{(3)} = \mathbb{E}_{ijk}^{(3)} - \left(\mathbb{E}_i^{(1)}\delta_{jk} + \mathbb{E}_j^{(1)}\delta_{ik} + \mathbb{E}_k^{(1)}\delta_{ij}\right) \tag{149}$$

$$A_{ijkl}^{(4)} = \mathbb{E}_{ijkl}^{(4)} - \left(\mathbb{E}_{ij}^{(2)}\delta_{kl} + \mathbb{E}_{ik}^{(2)}\delta_{jl} + \mathbb{E}_{il}^{(2)}\delta_{jk} + \mathbb{E}_{jk}^{(2)}\delta_{il} + \mathbb{E}_{jl}^{(2)}\delta_{ik} + \mathbb{E}_{kl}^{(2)}\delta_{ij}\right) + \mathbb{E}^{(0)}\delta_{ijkl} \tag{150}$$

Simplifying the above Hermite coefficients gives:

$$A^{(0)} = \rho \tag{151}$$

$$A_i^{(1)} = \rho \mathrm{v}_i \tag{152}$$

$$A_{ij}^{(2)} = \rho \mathrm{v}_i \mathrm{v}_j + \rho\tilde{\theta}\delta_{ij} \tag{153}$$

$$A_{ijk}^{(3)} = \rho \mathrm{v}_i \mathrm{v}_j \mathrm{v}_k + \rho\tilde{\theta}\left(\mathrm{v}_i\delta_{jk} + \mathrm{v}_j\delta_{ik} + \mathrm{v}_k\delta_{ij}\right) \tag{154}$$

$$A_{ijkl}^{(4)} = \rho \mathrm{v}_i \mathrm{v}_j \mathrm{v}_k \mathrm{v}_l + \rho\tilde{\theta}\left(\mathrm{v}_i\mathrm{v}_j\delta_{kl} + \mathrm{v}_i\mathrm{v}_k\delta_{jl} + \mathrm{v}_i\mathrm{v}_l\delta_{jk} + \mathrm{v}_j\mathrm{v}_k\delta_{il} + \mathrm{v}_j\mathrm{v}_l\delta_{ik} + \mathrm{v}_k\mathrm{v}_l\delta_{ij}\right) + \rho\tilde{\theta}^2\delta_{ijkl} \tag{155}$$

Where the variable $\tilde{\theta}$ has been introduced as $RT - 1$. Taking (137) and generating moments gives:

$$\int_{-\infty}^{+\infty} \mathcal{M}^* \boldsymbol{\xi}^m \,\mathrm{d}\boldsymbol{\xi} = \sum_{n=0}^{N} \frac{1}{n!} \boldsymbol{A}^{(n)} \cdot \int_{-\infty}^{+\infty} \omega \boldsymbol{H}^{(n)} \boldsymbol{\xi}^m \,\mathrm{d}\boldsymbol{\xi} \tag{156}$$

The operation $\boldsymbol{\xi}^m$ is a tensor product. The Hermite coefficients can be taken outside of the integral therefore the integrand becomes known functions which can be evaluated. The integrals on the right side of (156) become:

$$\int_{-\infty}^{+\infty} \omega \boldsymbol{H}^{(n)} \,\mathrm{d}\boldsymbol{\xi} = \begin{cases} 1 & n = 0 \\ 0 & n \neq 0 \end{cases} \tag{157}$$

$$\int_{-\infty}^{+\infty} \omega \boldsymbol{H}^{(n)} \xi_i \,\mathrm{d}\boldsymbol{\xi} = \begin{cases} \delta_{ij} & n = 1 \\ 0 & n \neq 1 \end{cases} \tag{158}$$

$$\int_{-\infty}^{+\infty} \omega \boldsymbol{H}^{(n)} \xi_i \xi_j \,\mathrm{d}\boldsymbol{\xi} = \begin{cases} \delta_{ij} & n = 0 \\ \delta^*_{ijkl} & n = 2 \\ 0 & n = 1, \quad n > 2 \end{cases} \tag{159}$$

$$\int_{-\infty}^{+\infty} \omega \boldsymbol{H}^{(n)} \xi_i \xi_j \xi_k \,\mathrm{d}\boldsymbol{\xi} = \begin{cases} \delta_{ijkl} & n = 1 \\ \delta^*_{ijklmn} & n = 3 \\ 0 & n = 0, \quad n = 2, \quad n > 3 \end{cases} \tag{160}$$

$$\int_{-\infty}^{+\infty} \omega \boldsymbol{H}^{(n)} \xi_i \xi_j \xi_k \xi_l \,\mathrm{d}\boldsymbol{\xi} = \begin{cases} \delta_{ijkl} & n = 0 \\ \delta^{**}_{ijklmn} & n = 2 \\ \delta^*_{ijklmnop} & n = 4 \\ 0 & \text{other} \end{cases} \tag{161}$$

Please see the appendix for the Kronecker delta identities. The truncated moments in (156) are evaluated as:

$$\mathbb{E}^*[\Xi^0] = \boldsymbol{A}^{(0)} \cdot \int_{-\infty}^{+\infty} \omega \boldsymbol{H}^{(0)} \, \mathrm{d}\boldsymbol{\xi} \tag{162}$$

$$\mathbb{E}^*[\Xi_i] = \boldsymbol{A}^{(1)} \cdot \int_{-\infty}^{+\infty} \omega \boldsymbol{H}^{(1)} \Xi_i \, \mathrm{d}\boldsymbol{\xi} \tag{163}$$

$$\mathbb{E}^*\left[\Xi_{ij}\right] = \boldsymbol{A}^{(0)} \cdot \int_{-\infty}^{+\infty} \omega \boldsymbol{H}^{(0)} \Xi_{ij} \, \mathrm{d}\boldsymbol{\xi} + \frac{1}{2} \boldsymbol{A}^{(2)} \cdot \int_{-\infty}^{+\infty} \omega \boldsymbol{H}^{(2)} \Xi_{ij} \, \mathrm{d}\boldsymbol{\xi} \tag{164}$$

$$\mathbb{E}^*\left[\Xi_{ijk}\right] = \boldsymbol{A}^{(1)} \cdot \int_{-\infty}^{+\infty} \omega \boldsymbol{H}^{(1)} \Xi_{ijk} \, \mathrm{d}\boldsymbol{\xi} + \frac{1}{6} \boldsymbol{A}^{(3)} \cdot \int_{-\infty}^{+\infty} \omega \boldsymbol{H}^{(3)} \Xi_{ijk} \, \mathrm{d}\boldsymbol{\xi} \tag{165}$$

$$\mathbb{E}^*\left[\Xi_{ijkl}\right] = \boldsymbol{A}^{(0)} \cdot \int_{-\infty}^{+\infty} \omega \boldsymbol{H}^{(0)} \Xi_{ijkl} \, \mathrm{d}\boldsymbol{\xi} + \frac{1}{2} \boldsymbol{A}^{(2)} \cdot \int_{-\infty}^{+\infty} \omega \boldsymbol{H}^{(2)} \Xi_{ijkl} \, \mathrm{d}\boldsymbol{\xi} + \frac{1}{24} \boldsymbol{A}^{(4)} \cdot \int_{-\infty}^{+\infty} \omega \boldsymbol{H}^{(4)} \Xi_{ijkl} \, \mathrm{d}\boldsymbol{\xi} \tag{166}$$

Thus yielding:

$$\mathbb{E}^*[\Xi^0] = A^{(0)} = \rho \quad \checkmark \tag{167}$$

$$\mathbb{E}^*[\Xi_i] = A_j^{(1)} \delta_{ij} = \rho \mathrm{v}_i \quad \checkmark \tag{168}$$

$$\mathbb{E}^*\left[\Xi_{ij}\right] = A^{(0)} \delta_{ij} + \frac{1}{2} A_{kl}^{(2)} \delta_{ijkl}^* = \rho \mathrm{v}_i \mathrm{v}_j + P \delta_{ij} \quad \checkmark \tag{169}$$

$$\mathbb{E}^*\left[\Xi_{ijk}\right] = A_l^{(1)} \delta_{ijkl} + \frac{1}{6} A_{lmn}^{(3)} \delta_{ijklmn}^* = \rho \mathrm{v}_i \mathrm{v}_j \mathrm{v}_k + P\left(\mathrm{v}_i \delta_{jk} + \mathrm{v}_j \delta_{ik} + \mathrm{v}_k \delta_{ij}\right) \quad \checkmark \tag{170}$$

The fourth order moment is evaluated using a MATLAB algorithm which is uploaded on GitHub. The fourth order primary moment of the fourth order Hermite expanded Maxwellian does not produce the original moment (22) however the partial moment (25) is fully recovered which is sufficient. The Maxwellian truncated at $N = 4$ and applying the Gauss-Hermite quadrature gives the discrete form:

$$\mathcal{M}_\alpha^* = w_\alpha \left( A^{(0)} H_\alpha^{(0)} + A_i^{(1)} H_{i,\alpha}^{(1)} + \frac{1}{2} A_{ij}^{(2)} H_{ij,\alpha}^{(2)} + \frac{1}{6} A_{ijk}^{(3)} H_{ijk,\alpha}^{(3)} + \frac{1}{24} A_{ijkl}^{(4)} H_{ijkl,\alpha}^{(4)} \right) \tag{171}$$

As a result, conditions for a discrete set of weights and lattices can be extracted from (156) by applying the Gauss-Hermite quadrature so that the integrals become finite sums:

$$\sum_{\alpha=1}^{Q} \mathcal{M}_\alpha^* \boldsymbol{\xi}_\alpha^m = \sum_{n=0}^{N} \frac{1}{n!} \boldsymbol{A}^{(n)} \cdot \sum_{\alpha=1}^{Q} w_\alpha \boldsymbol{H}_\alpha^{(n)} \boldsymbol{\xi}_\alpha^m \tag{172}$$

The complete set of required conditions are given as: (see appendix for reference)

Hermite Moments $\boldsymbol{H}^{(0)}$

$$\sum_{\alpha=1}^{Q} w_\alpha H_\alpha^{(0)} = 1 \qquad Q \geq 1 \tag{173}$$

Hermite Moments $\boldsymbol{H}^{(1)}$

$$\sum_{\alpha=1}^{Q} w_\alpha H_{i,\alpha}^{(1)} = 0 \qquad Q \geq 3 \tag{174}$$

$$\sum_{\alpha=1}^{Q} w_\alpha H_{i,\alpha}^{(1)} \xi_{j,\alpha} = \delta_{ij} \qquad Q \geq 3 \tag{175}$$

$$\sum_{\alpha=1}^{Q} w_\alpha H_{i,\alpha}^{(1)} \xi_{j,\alpha} \xi_{k,\alpha} = 0 \qquad Q \geq 5 \tag{176}$$

$$\sum_{\alpha=1}^{Q} w_\alpha H_{i,\alpha}^{(1)} \xi_{j,\alpha} \xi_{k,\alpha} \xi_{l,\alpha} = \delta_{ijkl} \qquad Q \geq 7 \tag{177}$$

$$\sum_{\alpha=1}^{Q} w_\alpha H_{i,\alpha}^{(1)} \xi_{j,\alpha} \xi_{k,\alpha} \xi_{l,\alpha} \xi_{m,\alpha} = 0 \qquad Q \geq 9 \tag{178}$$

Hermite Moments $\boldsymbol{H}^{(2)}$

$$\sum_{\alpha=1}^{Q} w_\alpha H_{ij,\alpha}^{(2)} = 0 \qquad Q \geq 5 \tag{179}$$

$$\sum_{\alpha=1}^{Q} w_\alpha H_{ij,\alpha}^{(2)} \xi_{k,\alpha} = 0 \qquad Q \geq 5 \tag{180}$$

$$\sum_{\alpha=1}^{Q} w_\alpha H_{ij,\alpha}^{(2)} \xi_{k,\alpha} \xi_{l,\alpha} = \delta_{ijkl}^{*} \qquad Q \geq 5 \tag{181}$$

$$\sum_{\alpha=1}^{Q} w_\alpha H_{ij,\alpha}^{(2)} \xi_{k,\alpha} \xi_{l,\alpha} \xi_{m,\alpha} = 0 \qquad Q \geq 7 \tag{182}$$

$$\sum_{\alpha=1}^{Q} w_\alpha H_{ij,\alpha}^{(2)} \xi_{k,\alpha} \xi_{l,\alpha} \xi_{m,\alpha} \xi_{n,\alpha} = \delta_{ijklmn}^{**} \qquad Q \geq 9 \tag{183}$$

Hermite Moments $\boldsymbol{H}^{(3)}$

$$\sum_{\alpha=1}^{Q} w_\alpha H_{ijk,\alpha}^{(3)} = 0 \qquad Q \geq 7 \tag{184}$$

$$\sum_{\alpha=1}^{Q} w_\alpha H_{ijk,\alpha}^{(3)} \xi_{l,\alpha} = 0 \qquad Q \geq 7 \tag{185}$$

$$\sum_{\alpha=1}^{Q} w_\alpha H_{ijk,\alpha}^{(3)} \xi_{l,\alpha} \xi_{m,\alpha} = 0 \qquad Q \geq 7 \tag{186}$$

$$\sum_{\alpha=1}^{Q} w_\alpha H_{ijk,\alpha}^{(3)} \xi_{l,\alpha} \xi_{m,\alpha} \xi_{n,\alpha} = \delta_{ijklmn}^{*} \qquad Q \geq 7 \tag{187}$$

$$\sum_{\alpha=1}^{Q} w_\alpha H_{ijk,\alpha}^{(3)} \xi_{l,\alpha} \xi_{m,\alpha} \xi_{n,\alpha} \xi_{o,\alpha} = 0 \qquad Q \geq 9 \tag{188}$$

Hermite Moments $\boldsymbol{H}^{(4)}$

$$\sum_{\alpha=1}^{Q} w_\alpha H_{ijkl,\alpha}^{(4)} = 0 \qquad Q \geq 9 \tag{189}$$

$$\sum_{\alpha=1}^{Q} w_\alpha H_{ijkl,\alpha}^{(4)} \xi_{m,\alpha} = 0 \qquad Q \geq 9 \tag{190}$$

$$\sum_{\alpha=1}^{Q} w_\alpha H_{ijkl,\alpha}^{(4)} \xi_{m,\alpha} \xi_{n,\alpha} = 0 \qquad Q \geq 9 \tag{191}$$

$$\sum_{\alpha=1}^{Q} w_\alpha H_{ijkl,\alpha}^{(4)} \xi_{m,\alpha} \xi_{n,\alpha} \xi_{o,\alpha} = 0 \qquad Q \geq 9 \tag{192}$$

$$\sum_{\alpha=1}^{Q} w_\alpha H_{ijkl,\alpha}^{(4)} \xi_{m,\alpha} \xi_{n,\alpha} \xi_{o,\alpha} \xi_{p,\alpha} = \delta_{ijklmnop}^{*} \qquad Q \geq 9 \tag{193}$$

Since the total energy distribution is a finite perturbation expansion of the Maxwellian, the established Gauss-Hermite velocity discretization procedure can be easily extended. The Hermite series expansion is:

$$\mathcal{E}^{*}(\boldsymbol{\xi}, \mathbf{v}, T, \rho) = \omega(\boldsymbol{\xi}) \sum_{m=0}^{M} \frac{1}{m!} \boldsymbol{B}^{(m)}(\mathbf{v}, T, \rho) \cdot \boldsymbol{H}^{(m)}(\boldsymbol{\xi}) \tag{194}$$

Where the total energy Hermite coefficients are defined as:

$$\boldsymbol{B}^{(m)}(\mathbf{v},T,\rho) = \int_{-\infty}^{+\infty} \mathcal{E}(\boldsymbol{\xi},\mathbf{v},T,\rho)\boldsymbol{H}^{(m)}(\boldsymbol{\xi})\,\mathrm{d}\boldsymbol{\xi} \tag{195}$$

Calculating from zeroth to second order gives:

$$B^{(0)} = \rho\mathrm{v}^2 + DP + aP + b \tag{196}$$

$$B_i^{(1)} = \rho\mathrm{v}^2\mathrm{v}_i + (D+2)P\mathrm{v}_i + aP\mathrm{v}_i \tag{197}$$

$$B_{ij}^{(2)} = \rho\mathrm{v}^2\left(\mathrm{v}_i\mathrm{v}_j - \delta_{ij}\right) + P\mathrm{v}^2\delta_{ij} + (D+4)P\mathrm{v}_i\mathrm{v}_j + P\left(D\tilde{\theta} + 2\theta\right)\delta_{ij} + aP\left(\mathrm{v}_i\mathrm{v}_j + \tilde{\theta}\delta_{ij}\right) + b\tilde{\theta}\delta_{ij} \tag{198}$$

Redefining in terms of the total scalar energy:

$$B^{(0)} = \rho E + b \tag{199}$$

$$B_i^{(1)} = (\rho E + 2P)\mathrm{v}_i \tag{200}$$

$$B_{ij}^{(2)} = (\rho E + 4P)\mathrm{v}_i\mathrm{v}_j + P(E + 2\theta)\delta_{ij} + b\theta\delta_{ij} - (\rho E + b)\delta_{ij} \tag{201}$$

This next step is not necessary but a velocity distribution vector can be defined as:

$$\mathcal{V}_i(\boldsymbol{\xi},\mathbf{v},T) = \frac{\xi_i}{(2\pi RT)^{D/2}}\mathrm{e}\left\{-\frac{1}{2RT}(\boldsymbol{\xi}-\mathbf{v})\cdot(\boldsymbol{\xi}-\mathbf{v})\right\} \tag{202}$$

Applying the Hermite series expansion gives the polynomial form:

$$\mathcal{V}_i^*(\boldsymbol{\xi},\mathbf{v},T) = \omega(\boldsymbol{\xi})\sum_{u=0}^{U}\frac{1}{u!}\boldsymbol{C}_i^{(u)}(\mathbf{v},T)\cdot\boldsymbol{H}^{(u)}(\boldsymbol{\xi}) \tag{203}$$

Where the Hermite coefficients are defined as:

$$\boldsymbol{C}_i^{(u)}(\mathbf{v},T) = \int_{-\infty}^{+\infty} \mathcal{V}_i(\boldsymbol{\xi},\mathbf{v},T)\boldsymbol{H}^{(u)}(\boldsymbol{\xi})\,\mathrm{d}\boldsymbol{\xi} \tag{204}$$

Calculating from zeroth to third order gives:

$$C_i^{(0)} = \mathrm{v}_i \tag{205}$$

$$C_{ij}^{(1)} = \mathrm{v}_i\mathrm{v}_j + \theta\delta_{ij} \tag{206}$$

$$C_{ijk}^{(2)} = \mathrm{v}_i\mathrm{v}_j\mathrm{v}_k + \theta\left(\mathrm{v}_i\delta_{jk} + \mathrm{v}_j\delta_{ik}\right) + \tilde{\theta}\mathrm{v}_k\delta_{ij} \tag{207}$$

$$C_{ijkl}^{(3)} = \mathrm{v}_i\mathrm{v}_j\mathrm{v}_k\mathrm{v}_l + \theta\big(\mathrm{v}_i\mathrm{v}_j\delta_{kl} + \mathrm{v}_i\mathrm{v}_k\delta_{jl} + \mathrm{v}_j\mathrm{v}_k\delta_{il}\big) + \tilde{\theta}\big(\mathrm{v}_i\mathrm{v}_l\delta_{jk} + \mathrm{v}_j\mathrm{v}_l\delta_{ik} + \mathrm{v}_k\mathrm{v}_l\delta_{ij}\big) + \theta\tilde{\theta}\delta_{ijkl} \tag{208}$$

Calculations for deriving the discrete set of weights and lattices at the Euler and N-S levels will next be shown.

Example: Euler Recovery

To recover the Euler equations (40-42), the solution in one dimension is referred to as the $D1Q7$ lattice where $\xi_\alpha = c\{0, \pm1, \pm2, \pm3\}$. The constant $D$ is the number of spatial dimensions and $Q$ is the number of lattices. Random velocity space is discretized into lattice vectors containing integer components multiplied by the lattice constant. The objective is to find $w_\alpha(c)$. In one dimension there are thirteen conditional equations however a lattice with only seven unknowns has been chosen being $w_\alpha \;\forall\; \alpha = 1 : Q$. The properties of the Hermite polynomials will be exploited to minimize the number of lattices and associated weights through the previously calculated integral in (157). A system of seven equations are needed to find the seven unknowns. The Hermite polynomials in one dimension from zeroth to sixth order are given as:

$$H^{(0)} = 1 \tag{209}$$

$$H^{(1)} = \xi \tag{210}$$

$$H^{(2)} = \xi^2 - 1 \tag{211}$$

$$H^{(3)} = \xi^3 - 3\xi \tag{212}$$

$$H^{(4)} = \xi^4 - 6\xi^2 + 3 \tag{213}$$

$$H^{(5)} = \xi^5 - 10\xi^3 + 15\xi \tag{214}$$

$$H^{(6)} = \xi^6 - 15\xi^4 + 45\xi^2 - 15 \tag{215}$$

Applying the Gauss-Hermite quadrature transformation to (157):

$$\int_{-\infty}^{+\infty} \omega \boldsymbol{H}^{(n)}\,\mathrm{d}\boldsymbol{\xi} = \sum_{\alpha=1}^{Q} w_\alpha \boldsymbol{H}_\alpha^{(n)} = \begin{cases} 1, & n = 0 \\ 0, & n \neq 0 \end{cases} \tag{216}$$

Applying (216) for $n = 0 : 6$ gives a system of seven equations with seven unknowns. A $Q \times Q$ matrix is formed as a function of $c$ denoted [M] and multiplied by the weights assembled into a $Q \times 1$ vector denoted $\{w\}$. This product is set equal to the resulting integral in (216) denoted {u} therefore giving the equation $[\mathrm{M}]\{w\} = \{\mathrm{u}\}$. The matrix [M] is calculated below where solutions are of the form $\{w\} = [\mathrm{M}]^{-1}\{\mathrm{u}\}$.

$\mathrm{M}_{11} = 1$
$\mathrm{M}_{21} = 0$
$\mathrm{M}_{31} = -1$
$\mathrm{M}_{41} = 0$
$\mathrm{M}_{51} = 3$
$\mathrm{M}_{61} = 0$
$\mathrm{M}_{71} = -15$

$\mathrm{M}_{12} = \mathrm{M}_{13} = 1$
$\mathrm{M}_{22} = \mathrm{M}_{23} = c$
$\mathrm{M}_{32} = \mathrm{M}_{33} = c^2 - 1$
$\mathrm{M}_{42} = \mathrm{M}_{43} = c^3 - 3c$
$\mathrm{M}_{52} = \mathrm{M}_{53} = c^4 - 6c^2 + 3$
$\mathrm{M}_{62} = \mathrm{M}_{63} = c^5 - 10c^3 + 15c$
$\mathrm{M}_{72} = \mathrm{M}_{73} = c^6 - 15c^4 + 45c^2 - 15$

$\mathrm{M}_{14} = +\mathrm{M}_{15} = 1$
$\mathrm{M}_{24} = -\mathrm{M}_{25} = 2c$

$\mathrm{M}_{16} = +\mathrm{M}_{17} = 1$
$\mathrm{M}_{26} = -\mathrm{M}_{27} = 3c$

$$\mathrm{M}_{34} = +\mathrm{M}_{35} = 4c^2 - 1$$
$$\mathrm{M}_{44} = -\mathrm{M}_{45} = 8c^3 - 6c$$
$$\mathrm{M}_{54} = +\mathrm{M}_{55} = 16c^4 - 24c^2 + 3$$
$$\mathrm{M}_{64} = -\mathrm{M}_{65} = 32c^5 - 80c^3 + 30c$$
$$\mathrm{M}_{74} = +\mathrm{M}_{75} = 64c^6 - 240c^4 + 180c^2 - 15$$

$$\mathrm{M}_{36} = +\mathrm{M}_{37} = 9c^2 - 1$$
$$\mathrm{M}_{46} = -\mathrm{M}_{47} = 27c^3 - 9c$$
$$\mathrm{M}_{56} = +\mathrm{M}_{57} = 81c^4 - 54c^2 + 3$$
$$\mathrm{M}_{66} = -\mathrm{M}_{67} = 243c^5 - 270c^3 + 45c$$
$$\mathrm{M}_{76} = +\mathrm{M}_{77} = 729c^6 - 121c^4 + 405c^2 - 15$$

Below are the calculated weights as a function of the lattice constant:

$$\begin{aligned} w_{1:1} &= (36c^6 - 49c^4 + 42c^2 - 15)/(36c^6) \\ w_{2:3} &= (12c^4 - 13c^2 + 5)/(16c^6) \\ w_{4:5} &= (-3c^4 + 10c^2 - 5)/(40c^6) \\ w_{6:7} &= (4c^4 - 15c^2 + 15)/(720c^6) \end{aligned} \tag{217}$$

Lastly (217) is fed into (173-193) $\forall\, Q \leq 7$ and all required conditions are satisfied. Trying the $D1Q5$ lattice where $\xi_\alpha = c\{0, \pm 1, \pm 2\}$ results in the final condition (187) giving $4c^4 - 15c^2 + 15 = 0$ which does not have real solutions therefore the $D1Q7$ lattice is minimal in one spatial dimension to recover the Eulerian transport equations exactly. The integral in (138) therefore is not approximated but shown to be exactly equal to a finite sum where the discrete set of weights are functions of $\{c \in \mathbb{R} | c \neq 0\}$. Although a continuous energy spectrum was assumed, hence the integrals in (15), the mathematics lead to a finite sum of the Maxwellian over a finite number of velocity states. Extending the solution to two and three dimensions can be done with ease due to the orthogonality of the Hermite polynomials. For the lattices, the number of ordered arrangements with replacement is given by the equation $Q = 7^D$ which leads to the $D2Q49$ and $D3Q343$. The weight at each lattice grid point can be written as the tensor product $w_\alpha w_\beta$ in two dimensions and $w_\alpha w_\beta w_\gamma$ in three dimensions which satisfy (173-193) $\forall\, Q \leq 7$.

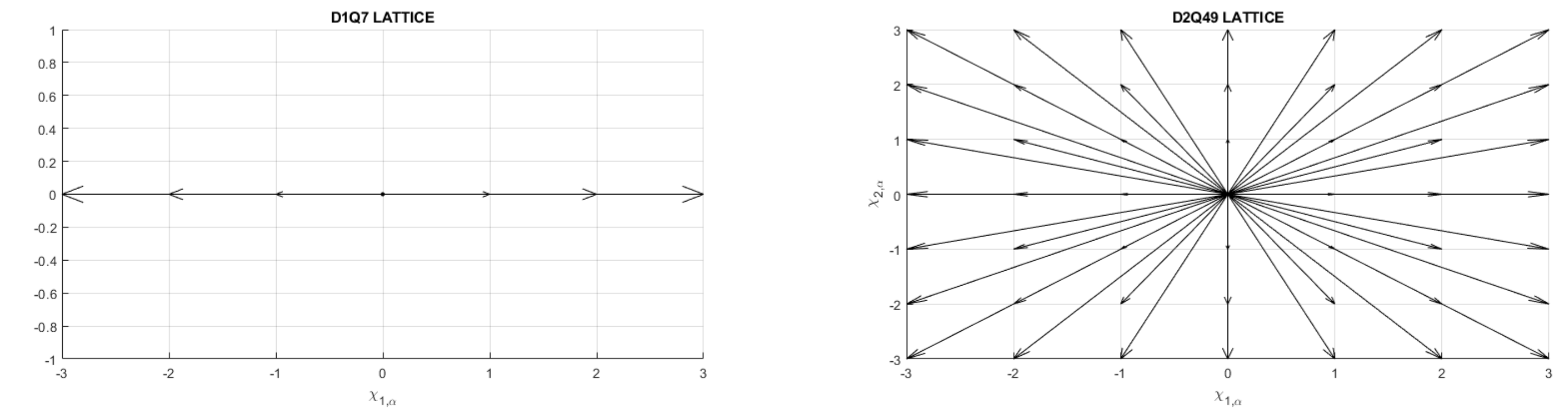

Figure #3 – Standard Complete Lattices $Q = 7^D$

In general, the equation $Q = (2N + 1)^D$ gives the required number of lattices based on the expansion number. To reduce the number of lattices, approximation methods can be used, see [8,9]. Please see below for examples:

<u>$D2Q17$</u> (see quadrature $E^7_{2,17}$ in [8]) $c^2 = \left(125 + 5\sqrt{193}\right)/72$

$$w_{1:4} = \left(3355 - 91\sqrt{193}\right)/18000 \qquad w_{5:8} = \left(655 + 17\sqrt{193}\right)/27000$$
$$w_{9:12} = \left(685 - 49\sqrt{193}\right)/54000 \qquad w_{13:16} = \left(1445 - 101\sqrt{193}\right)/162000$$
$$w_{17} = \left(575 + 193\sqrt{193}\right)/8100$$

$$\xi_{1,1:4} = c\cos\{(\pi/2)(\alpha - 1)\} \qquad \xi_{2,1:4} = c\sin\{(\pi/2)(\alpha - 1)\}$$
$$\xi_{1,5:8} = \sqrt{2}c\cos\{(\pi/2)(\alpha - 1) + \pi/4\} \qquad \xi_{2,5:8} = \sqrt{2}c\sin\{(\pi/2)(\alpha - 1) + \pi/4\}$$

$\xi_{1,9:12} = 2\sqrt{2}c\cos\{(\pi/2)(\alpha-1)+\pi/4\}$ $\quad$ $\xi_{2,9:12} = 2\sqrt{2}c\sin\{(\pi/2)(\alpha-1)+\pi/4\}$
$\xi_{1,13:16} = 3c\cos\{(\pi/2)(\alpha-1)\}$ $\quad$ $\xi_{2,13:16} = 3c\sin\{(\pi/2)(\alpha-1)\}$
$\xi_{1,17} = 0$ $\quad$ $\xi_{2,17} = 0$

*D3Q39* (see quadrature $E^7_{3,39}$ in [8]) $c^2 = 3/2$

$$
\begin{aligned}
\xi_{i,1:3} &= +c\delta_{i,\alpha} \\
\xi_{i,4:6} &= -c\delta_{i,\alpha-3} \\
\xi_{1,7:14} &= \sqrt{2}c\cos\{(\pi/2)(\alpha-1)+\pi/4\} \\
\xi_{2,7:14} &= \sqrt{2}c\sin\{(\pi/2)(\alpha-1)+\pi/4\} \\
\xi_{3,7:14} &= \pm c\left(+ce_{3,7:10}, -ce_{3,11:14}\right) \\
\xi_{i,15:17} &= +2c\delta_{i,\alpha-14} \\
\xi_{i,18:20} &= -2c\delta_{i,\alpha-17} \\
\xi_{1,21:28} &= \xi_{2,29:32} = 2\sqrt{2}c\cos\{(\pi/2)(\alpha-1)+\pi/4\} \\
\xi_{2,21:24} &= \xi_{3,25:32} = 2\sqrt{2}c\sin\{(\pi/2)(\alpha-1)+\pi/4\} \\
\xi_{3,21:24} &= \xi_{2,25:28} = \xi_{1,29:32} = 0 \\
\xi_{i,33:35} &= +3c\delta_{i,\alpha-32} \\
\xi_{i,36:38} &= -3c\delta_{i,\alpha-35} \\
\xi_{i,39} &= 0
\end{aligned}
$$

$$
\begin{aligned}
w_{1:6} &= 1/12 \\
w_{7:14} &= 1/27 \\
w_{15:20} &= 2/135 \\
w_{21:32} &= 1/432 \\
w_{33:38} &= 1/1620 \\
w_{39} &= 1/12
\end{aligned}
$$

It should also be noted that the above method allows for alternate lattices. For example, a sufficient lattice scheme can be $\xi_\alpha = \chi_\alpha/\tau$ where $\chi_\alpha$ can take on number types other than integers and be nonsymmetric. To satisfy the constraints, a new set of weights can be derived. In conclusion, the integrals in the Eulerian system (48-50) have been replaced with finite sums:

0th Moment: $$\frac{\partial}{\partial t}\sum_{\alpha=1}^{Q}\mathcal{M}^*_\alpha = -\frac{\partial}{\partial x_i}\sum_{\alpha=1}^{Q}\mathcal{M}^*_\alpha\xi_{i,\alpha} \tag{218}$$

1st Moment: $$\frac{\partial}{\partial t}\sum_{\alpha=1}^{Q}\mathcal{M}^*_\alpha\xi_{i,\alpha} = -\frac{\partial}{\partial x_j}\sum_{\alpha=1}^{Q}\mathcal{M}^*_\alpha\xi_{i,\alpha}\xi_{j,\alpha} \tag{219}$$

2nd Scalar Moment: $$\frac{\partial}{\partial t}\sum_{\alpha=1}^{Q}\mathcal{M}^*_\alpha\xi^2_\alpha = -\frac{\partial}{\partial x_i}\sum_{\alpha=1}^{Q}\mathcal{M}^*_\alpha\xi^2_\alpha\xi_{i,\alpha} \tag{220}$$

Example: Navier-Stokes Recovery

In one dimension, the $D1Q7$ lattice satisfies all conditions except (193) which gives the equation:

$$36c^6 - 147c^4 + 210c^2 - 81 = 24 \tag{221}$$

To satisfy this final condition, real solutions exist for the lattice constant:

$$c = \frac{\sqrt{\left(469+252\sqrt{30}\right)^{2/3} + 49\left(469+252\sqrt{30}\right)^{1/3} - 119}}{6\left(469+252\sqrt{30}\right)^{1/6}} \approx \pm 1.196979771 \tag{222}$$

The above value matches the computed value in [8,9] (please note with courtesy there is a typographical error). If one wants to avoid this restriction, the $D1Q9$ lattice can be utilized where $\xi_\alpha = c\{0, \pm 1, \pm 2, \pm 3, \pm 4\}$. The associated weights are then derived as:

$$\begin{aligned} w_{1:1} &= (576c^8 - 820c^6 + 819c^4 - 450c^2 + 105)/(576c^8) \\ w_{2:3} &= (192c^6 - 244c^4 + 145c^2 - 35)/(240c^8) \\ w_{4:5} &= (-48c^6 + 169c^4 - 130c^2 + 35)/(480c^8) \\ w_{6:7} &= (64c^6 - 252c^4 + 315c^2 - 105)/(5040c^8) \\ w_{8:9} &= (-12c^6 + 49c^4 - 70c^2 + 35)/(13440c^8) \end{aligned} \quad (223)$$

In two dimensions exists the $D2Q81$ lattice and in three the $D3Q729$. It will also be noted for energy conservation, in one dimension, the $D1Q5$ scheme is minimally sufficient to be applied to the discrete total energy distribution (194). The weights are derived as:

$$w_{1:1} = (4c^4 - 5c^2 + 3)/(4c^4), \qquad w_{2:3} = (4c^2 - 3)/(6c^4), \qquad w_{4:5} = (-c^2 + 3)/(24c^4) \quad (224)$$

In two dimensions exists the $D2Q25$ lattice and in three the $D3Q125$. For $c = 1$, the derived weights are summarized in one dimension:

| $\underline{D1Q9}$ | $\underline{D1Q7}$ | $\underline{D1Q5}$ | |
|---|---|---|---|
| $w_{1:1} = 115/288$ | $w_{1:1} = 7/18$ | $w_{1:1} = 1/2$ | |
| $w_{2:3} = 29/120$ | $w_{2:3} = 1/4$ | $w_{2:3} = 1/6$ | (225) |
| $w_{4:5} = 13/240$ | $w_{4:5} = 1/20$ | $w_{4:5} = 1/12$ | |
| $w_{6:7} = 11/2520$ | $w_{6:7} = 1/180$ | | |
| $w_{8:9} = 1/6720$ | | | |

At the $N = 4$ level ($D1Q9$), an approximated lower order lattice in two dimensions is given below where the lattice constant $c$ is the computed value in (222).

$\underline{D2Q37}$, see [9,16]

$w(1:4) = 0.23315066913235250228650$ $\quad$ $w(17:20) = 0.00101193759267357547541$
$w(5:8) = 0.10730609154221900241246$ $\quad$ $w(21:28) = 0.00024530102775771734547$
$w(9:12) = 0.05766785988879488203006$ $\quad$ $w(29:36) = 0.00535304900051377523273$
$w(13:16) = 0.01420821615845075026469$ $\quad$ $w(37) = 0.00028341425299419821740$

| $\alpha$ | $\xi_{x\alpha}$ | $\xi_{y\alpha}$ |
|---|---|---|
| $1:4$ | $c\cos\{(\pi/2)(\alpha-1)\}$ | $c\sin\{(\pi/2)(\alpha-1)\}$ |
| $5:8$ | $\sqrt{2}c\cos\{(\pi/2)(\alpha-5)+\pi/4\}$ | $\sqrt{2}c\sin\{(\pi/2)(\alpha-5)+\pi/4\}$ |
| $9:12$ | $2c\cos\{(\pi/2)(\alpha-1)\}$ | $2c\sin\{(\pi/2)(\alpha-1)\}$ |
| $13:16$ | $2\sqrt{2}c\cos\{(\pi/2)(\alpha-5)+\pi/4\}$ | $2\sqrt{2}c\sin\{(\pi/2)(\alpha-5)+\pi/4\}$ |
| $17:20$ | $3c\cos\{(\pi/2)(\alpha-1)\}$ | $3c\sin\{(\pi/2)(\alpha-1)\}$ |
| $21:24$ | $\sqrt{5}c\cos\{(\pi/2)(\alpha-21)+\tan^{-1}1/2\}$ | $\sqrt{5}c\sin\{(\pi/2)(\alpha-21)+\tan^{-1}1/2\}$ |
| $25:28$ | $\sqrt{5}c\cos\{(\pi/2)(\alpha-25)+\tan^{-1}2\}$ | $\sqrt{5}c\sin\{(\pi/2)(\alpha-25)+\tan^{-1}2\}$ |
| $29:32$ | $\sqrt{10}c\cos\{(\pi/2)(\alpha-29)+\tan^{-1}1/3\}$ | $\sqrt{10}c\sin\{(\pi/2)(\alpha-29)+\tan^{-1}1/3\}$ |
| $33:36$ | $\sqrt{10}c\cos\{(\pi/2)(\alpha-33)+\tan^{-1}3\}$ | $\sqrt{10}c\sin\{(\pi/2)(\alpha-33)+\tan^{-1}3\}$ |
| $37$ | $0$ | $0$ |

# FOURIER SERIES ANALYSIS

In this next section, the discrete Maxwellian will be analyzed. Considering a steady state in a uniform space, meaning $\rho$, $\mathbf{v}$, and $T$ are continuous constants throughout the domain of $\boldsymbol{x}$, the discrete Maxwellian can be represented as a set of piecewise functions which become periodic in space. The discrete lattice was previously written as $\boldsymbol{\xi}_\alpha = c\boldsymbol{e}_\alpha$. The lattice constant $c$ is now recognized to become $1/\tau$ which is a temporal frequency. The integer vector $\boldsymbol{e}_\alpha$ can then be written as $\boldsymbol{\chi}_\alpha$ therefore giving the new discrete velocity equation $\boldsymbol{\xi}_\alpha = \boldsymbol{\chi}_\alpha/\tau$. In a continuous sense, the velocity differential becomes $\mathrm{d}\boldsymbol{\xi} = \mathrm{d}\boldsymbol{\chi}/\tau^D$ therefore further realizing the moments purely depend on $\boldsymbol{\chi}$ and the temporal frequency was always constant. The moments can be rewritten as:

$$\int_{-\infty}^{+\infty} \mathcal{M}(\boldsymbol{\xi})\boldsymbol{\xi}^m \,\mathrm{d}\boldsymbol{\xi} = \frac{1}{\tau^{D+m}} \int_{-\infty}^{+\infty} \mathcal{M}(\boldsymbol{\xi})\boldsymbol{\chi}^m \,\mathrm{d}\boldsymbol{\chi} \;\Rightarrow\; \sum_{\alpha=1}^{Q} \mathcal{M}_\alpha^* \boldsymbol{\xi}_\alpha^m = \frac{1}{\tau^m} \sum_{\alpha=1}^{Q} \mathcal{M}_\alpha^* \boldsymbol{\chi}_\alpha^m \tag{226}$$

In one dimension, for example the $D1Q7$, there are seven discrete distribution values therefore resulting in seven periodic piecewise functions throughout space. Each $\mathcal{M}_\alpha^*$ is assumed to occupy the space halfway to the left and right of its corresponding spatial position therefore occupying one unit of total space. Taking a leap of faith, in a quantum mechanical sense, each discrete Maxwellian occupies a potential well of unit length. A Fourier series representation will now be utilized thus revealing seven unique periodic wave functions:

$$\psi_\alpha(x) = \frac{1}{Q} \int_{\mathcal{A}_\alpha}^{\mathcal{B}_\alpha} \mathcal{M}_\alpha^* \,\mathrm{d}x + \sum_{n=1}^{\infty} \mathrm{A}_{\alpha,n} \cos 2\pi n x/Q + \sum_{n=1}^{\infty} \mathrm{B}_{\alpha,n} \sin 2\pi n x/Q \tag{227}$$

Where the Fourier coefficients are given as:

$$\mathrm{A}_{\alpha,n} = \frac{2}{Q} \int_{\mathcal{A}_\alpha}^{\mathcal{B}_\alpha} \mathcal{M}_\alpha^* \cos 2\pi n x/Q \,\mathrm{d}x = 2\mathcal{M}_\alpha^*/(n\pi) \sin n\pi/Q \,(2\cos^2 n\pi\chi_\alpha/Q - 1) \tag{228}$$

$$\mathrm{B}_{\alpha,n} = \frac{2}{Q} \int_{\mathcal{A}_\alpha}^{\mathcal{B}_\alpha} \mathcal{M}_\alpha^* \sin 2\pi n x/Q \,\mathrm{d}x = 4\mathcal{M}_\alpha^*/(n\pi) \sin n\pi/Q \sin n\pi\chi_\alpha/Q \cos n\pi\chi_\alpha/Q \tag{229}$$

The halfway interval length $L$ is equal to $Q/2$ where $Q$ is the distance between common grid points. The integration bounds represent the half steps between each lattice node where the total distance between nodes is one unit therefore the bounds are $\chi_\alpha \pm 1/2$. The moments of the wave functions will be referred to as the wave moments denoted $\Psi_m(x)$. Due to the periodic nature of the discrete distributions, the wave moments are continuous functions in space that contain the discrete information of the lattices. For example, the zeroth moment wave contains all values of the discrete Maxwellian in one function dependent upon space. The zeroth moment wave is the representation function of all values of each discrete distribution which is dependent upon various distances $\chi_\alpha$ from $x$. The wave moments are given as:

$$\Psi_m(x) = \sum_{\alpha=1}^{Q} \psi_\alpha(x)\xi_\alpha^m = \frac{1}{\tau^m} \sum_{\alpha=1}^{Q} \psi_\alpha(x)\chi_\alpha^m \tag{230}$$

To obtain the physical moments at a point in space along $\Psi_m(x)$, the wave moments will be summed over each discrete lattice distance:

$$\mathbb{E}[\xi_\alpha^m] = \sum_{\beta=1}^{Q} \Psi_m(x + \chi_\beta) = \frac{1}{\tau^m} \sum_{\beta=1}^{Q} \sum_{\alpha=1}^{Q} \psi_\alpha(x + \chi_\beta)\chi_\alpha^m \tag{231}$$

The Fourier series representation is simply a unique feature of the velocity discretization. It does however allow for a deeper interpretation of the discrete lattices. Expanding (230) gives the form:

$$\Psi_m(x)\tau^m = \psi_1(x)\chi_1^m + \psi_2(x)\chi_2^m + \cdots + \psi_Q(x)\chi_Q^m \tag{232}$$

Comparing with the definition of the general integral:

$$F(x) = \lim_{n\to\infty} \sum_{i=1}^{n} f(x_i)\Delta x_i = f(x_1)\Delta x_1 + f(x_2)\Delta x_2 + \cdots + f(x_n)\Delta x_n \tag{233}$$

The difference is that $\psi_\alpha$ is a set of functions where $f$ is a single function evaluated at $x_i$. Expanding (231):

$$\begin{aligned} \mathbb{E}[\chi_\alpha^m] = \Psi_m(x + \chi_1) + \Psi_m(x + \chi_2) + \cdots + \Psi_m(x + \chi_Q) = \\ \psi_1(x + \chi_1)\chi_1^m + \psi_2(x + \chi_1)\chi_2^m + \cdots + \psi_Q(x + \chi_1)\chi_Q^m + \\ \psi_1(x + \chi_2)\chi_1^m + \psi_2(x + \chi_2)\chi_2^m + \cdots + \psi_Q(x + \chi_2)\chi_Q^m + \\ \psi_1(x + \chi_Q)\chi_1^m + \psi_2(x + \chi_Q)\chi_2^m + \cdots + \psi_Q(x + \chi_Q)\chi_Q^m \end{aligned} \tag{234}$$

The stepping delta element in the genera integral will be simplified to a constant by setting $\Delta x_i = (i - 1)\Delta x$ which gives $x_i = x + (i - 1)\Delta x$ therefore the integral becomes:

$$F(x) = \Delta x f(x + \Delta x) + 2\Delta x f(x + 2\Delta x) + \cdots + (n - 1)\Delta x f(x + (n - 1)\Delta x) \tag{235}$$

The integral is of course an infinite sum where (231) has been shown to showcase the transformation to a discrete sum though the use of multiple functions. To end this section, a PDE satisfied by (227) will be searched for. Beginning with the discrete freestream Boltzmann equation in one dimension yields an exponential propagation solution where the Fourier series involves trigonometric functions, particularly the sine and cosine functions. To transform this PDE, the stepping ratio $\chi_\alpha/\tau$ will be converted into the operator $-(\chi_\alpha/\tau)\,\chi_\alpha\partial/\partial x$:

$$\frac{\partial}{\partial t}\Psi_\alpha(x,t) = \frac{\chi_\alpha^2}{\tau}\frac{\partial^2}{\partial x^2}\Psi_\alpha(x,t) \tag{236}$$

In a uniform space, homogeneous boundary conditions exist therefore it is valid to apply the method of separation of variables giving $\Psi_\alpha(x,t) = \psi_\alpha(x)\varphi_\alpha(t)$. The spatial differential eigenvalue problem is obtained as $\chi_\alpha^2\psi_\alpha'' = -\psi_\alpha\lambda^2$ which produces solutions of the form:

$$\psi_\alpha(x) = \mathrm{A}_\alpha \cos \lambda x/\chi_\alpha + \mathrm{B}_\alpha \sin \lambda x/\chi_\alpha + \mathrm{C}_\alpha \tag{237}$$

Combing the method of superposition with the homogeneous boundary conditions obtained from the discrete Maxwellian allows one to conclude $\lambda/\chi_\alpha = 2\pi n/Q$, see [6] (Ch. 2). As a result, the solution can be written as a sum of solutions:

$$\psi_\alpha(x) = \sum_{n=1}^{\infty} \mathrm{A}_{\alpha,n} \cos n\pi x/L + \sum_{n=1}^{\infty} \mathrm{B}_{\alpha,n} \sin n\pi x/L + \mathrm{C}_\alpha \tag{238}$$

Which reveals a Fourier series where $Q = 2L$. For the sake of completeness, the temporal ODE from the method of separation of variables becomes $\tau\dot{\varphi}_\alpha = -\varphi_\alpha\lambda^2$ which generates the solution $\varphi_\alpha = \exp\{-\lambda^2 t/\tau\}$. In conclusion, (227) is a solution to the derived PDE (236). To end this section, please see the figure below which plots the cosine and sine Fourier series representation of the discrete Maxwellian.

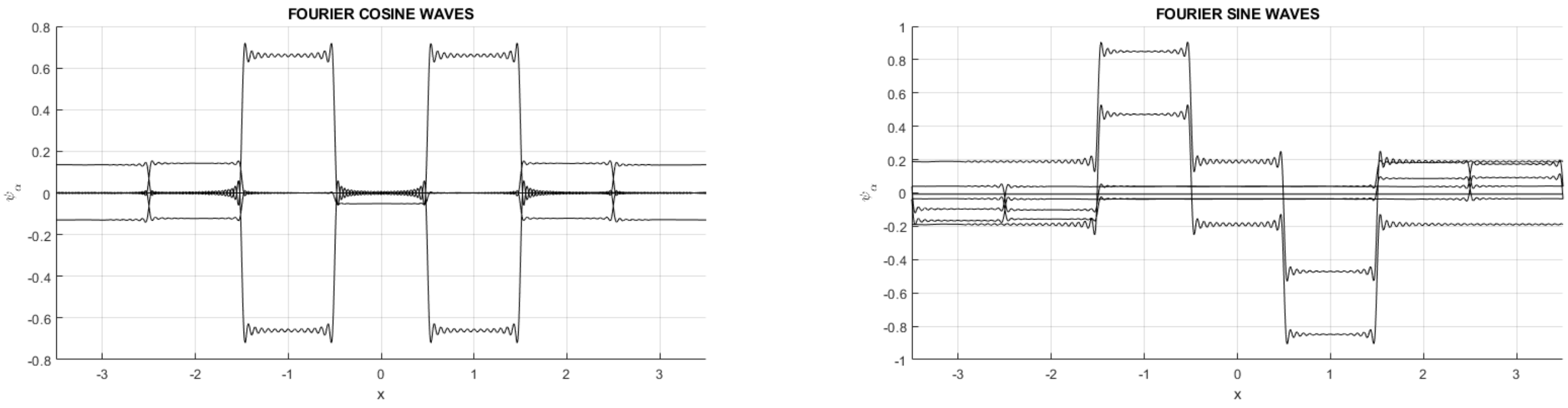

Figure #4 – Fourier Series Plots ($n = 100$)

In summary, a deeper theoretical understanding of the discrete Maxwellian has been obtained through application of the Fourier series. This concludes the theory behind this work which will be concluded in the next section through a brief elementary application.

## A SIMPLE NUMERICAL SCHEME

A basic computational procedure at the Euler level will now be investigated. The transport equations have been reformed in terms of discrete distribution functions below. The time and spatial derivatives require resolving for numerical analysis. To recap, at the Euler level, for a compressible inviscid fluid, the distribution evolutions are:

$$\frac{\partial}{\partial t}\mathcal{M}_\alpha^* = -\frac{\partial}{\partial x_i}\mathcal{M}_\alpha^*\xi_{i,\alpha}, \qquad \frac{\partial}{\partial t}\mathcal{M}_\alpha^*\xi_{i,\alpha} = -\frac{\partial}{\partial x_j}\mathcal{M}_\alpha^*\xi_{i,\alpha}\xi_{j,\alpha}, \qquad \frac{\partial}{\partial t}\mathcal{E}_\alpha^* = -\frac{\partial}{\partial x_i}\mathcal{E}_\alpha^*\xi_{i,\alpha} \tag{239}$$

The distribution evolution equation to be numerically evaluated is then summarized:

$$\frac{\partial}{\partial t}\mathbb{f}_\alpha^{\mathbb{m}}(\boldsymbol{x},t) = -\xi_{i,\alpha}\frac{\partial}{\partial x_i}\mathbb{f}_\alpha^{\mathbb{m}}(\boldsymbol{x},t) \tag{240}$$

Applying a first order Euler forward finite difference scheme to the time derivative in (240) allows for the numerical evolution of the summational invariants. Unfolding (111-113) or equivalently (239) gives:

$$\mathcal{M}_{I,\alpha}^{t+1} = \mathcal{M}_{I,\alpha}^{t} - \tau\nabla_k(\mathcal{M}\xi_k)_{I,\alpha}^t \;\Rightarrow\; \rho_I^{t+1} \tag{241}$$

$$(\mathcal{M}\xi_i)_{I,\alpha}^{t+1} = (\mathcal{M}\xi_i)_{I,\alpha}^{t} - \tau\nabla_j\big(\mathcal{M}\xi_i\xi_j\big)_{I,\alpha}^t \;\Rightarrow\; p_{i,I}^{t+1} \;\Rightarrow\; \mathrm{v}_{i,I}^{t+1} = p_{i,I}^{t+1}/\rho_I^{t+1} \tag{242}$$

$$\mathcal{E}_{I,\alpha}^{t+1} = \mathcal{E}_{I,\alpha}^{t} - \tau\nabla_k(\mathcal{E}\xi_k)_{I,\alpha}^t \;\Rightarrow\; e_I^{t+1} + b \;\Rightarrow\; E_I^{t+1} = e_I^{t+1}/\rho_I^{t+1} \tag{243}$$

The temperature and pressure will be calculated via:

$$T_I^{t+1} = \left(E_I^{t+1} - \mathrm{v}_{k,I}^{t+1}\mathrm{v}_{k,I}^{t+1}\right)/(AR) \;\;\Rightarrow\;\; P_I^{t+1} = \rho_I^{t+1} R T_I^{t+1} \tag{244}$$

The order of solving must start from the zeroth moment level then sequentially to the $m+1$ moment level. First, in (241) the discrete distribution is temporally advanced so that the new density can be calculated. Second, (242) is executed to find the new momentum so that the new velocity can be found. The first and second steps can be combined into one or (242) can be rewritten in terms of the velocity distribution vector (202). Again, this is not a necessary step and can be avoided to save algorithm execution time. The final step (243) allows for the calculation of the total energy density from which the temperature and pressure can be found. The discrete distributions will then need to be updated from the new summational invariants:

$$\mathcal{M}_{I,\alpha}^{t+1} = w_\alpha^A \sum_{n=0}^{N} \frac{1}{n!} \boldsymbol{H}_\alpha^{(n)} \cdot \boldsymbol{A}_I^{t+1(n)}, \qquad \mathcal{E}_{I,\alpha}^{t+1} = w_\alpha^B \sum_{m=0}^{M} \frac{1}{m!} \boldsymbol{H}_\alpha^{(m)} \cdot \boldsymbol{B}_I^{t+1(m)} \tag{245}$$

Applying a first order Euler forward finite difference scheme in time to (240) and a first order Euler backward scheme in space gives the following derivative transformations respectively:

$$\frac{\partial}{\partial t}\mathbb{f}_\alpha^{\mathbb{m}}(\boldsymbol{x},t) = \frac{\mathbb{f}_\alpha^{\mathbb{m}}(\boldsymbol{x},t+\tau) - \mathbb{f}_\alpha^{\mathbb{m}}(\boldsymbol{x},t)}{\tau}, \qquad \frac{\partial}{\partial x_i}\mathbb{f}_\alpha^{\mathbb{m}}(\boldsymbol{x},t) = \frac{\mathbb{f}_\alpha^{\mathbb{m}}(\boldsymbol{x},t) - \mathbb{f}_\alpha^{\mathbb{m}}(\boldsymbol{x}-\boldsymbol{\chi}_\alpha,t)}{\chi_{i,\alpha}} \tag{246}$$

After setting $\boldsymbol{\xi}_\alpha = \boldsymbol{\chi}_\alpha/\tau$ and applying the above schemes to (240) gives the simple and elegant equation:

$$\mathbb{f}_\alpha^{\mathbb{m}}(\boldsymbol{x},t+\tau) = \mathbb{f}_\alpha^{\mathbb{m}}(\boldsymbol{x}-\boldsymbol{\chi}_\alpha,t) \tag{247}$$

Where the Courant-Friedrichs-Lewy (CFL) number is exactly one on a uniform grid. The above discretized evolution is the equivalent of a first order upwind finite difference scheme. To test the algorithm, an example problem in one dimension will first be simulated. For numerical stability purposes, the Maxwellian will be expanded to $N=4$ and the total energy distribution to $M=2$. A single gas will be separated by a barrier leaving a left and right region. The initial conditions are given as:

$$\rho_L = P_L = 1, \qquad \rho_R = P_R = 1/2, \qquad \mathrm{v}_L = \mathrm{v}_R = 0 \tag{248}$$

The $D1Q9$ and $D1Q5$ Gauss-Hermite quadratures will be applied to the Maxwellian and total energy distributions respectively. The discrete velocity lattice will be defined as $\xi_\alpha = \chi_\alpha/\tau$ where $\chi_\alpha$ can take on non-integer values. The discrete integer spatial steps $\chi_\alpha^* = \{0, \pm 1, \pm 2, \cdots\}$ will be linked to the uniform grid spacing $\Delta x$ through the equation $\chi_\alpha = \chi_\alpha^* \Delta x$ where $\chi_\alpha^*$ is a nondimensional quantity. As a result, the weights are now dependent upon both $\Delta x$ and $\tau$. The time step is set equal to $\tau$ and the lattice constant is set to $c = 1$. The spatial position $x$ will be uniformly discretized and will now be referred to as the position $I$ of a vector containing all discrete spatial points. The adiabatic index is $\Gamma = 7/5$ and the boundary conditions are set equal to the initial conditions. Please see the next page plotting the results. The numerical solution (black dots) is compared with an analytical solution (dashed red line) taken from [15]. The expansion fan appears to be captured exactly however there are slight deviations between results in the shock zone regions. If the difference between left and right regions in the initial conditions is increased, (247) becomes unstable even with higher order expansions and quadratures. To prove validity in higher dimensions, a similar benchmark problem will be analyzed. In two dimensions, the shock tube becomes a shock box with inner and outer regions matching the initial conditions from the previous. See [13] for comparison of a similar simulation. The solutions appear to be smooth and reliable however limited to the difference between initial conditions where more measures will need to be taken.

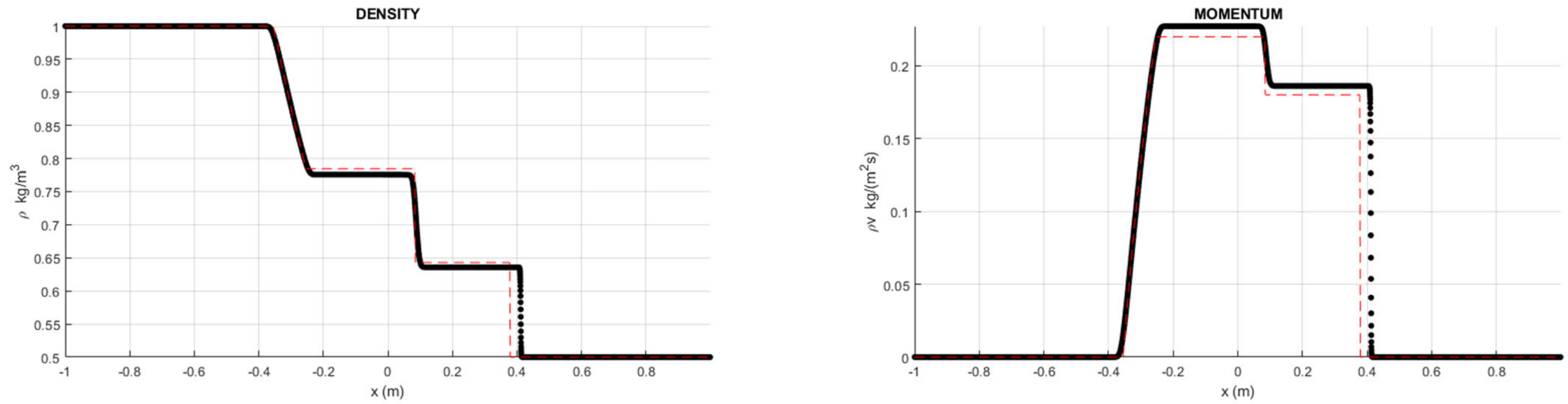

Figure #5 – Density & Momentum ($t = 0.3$ s)

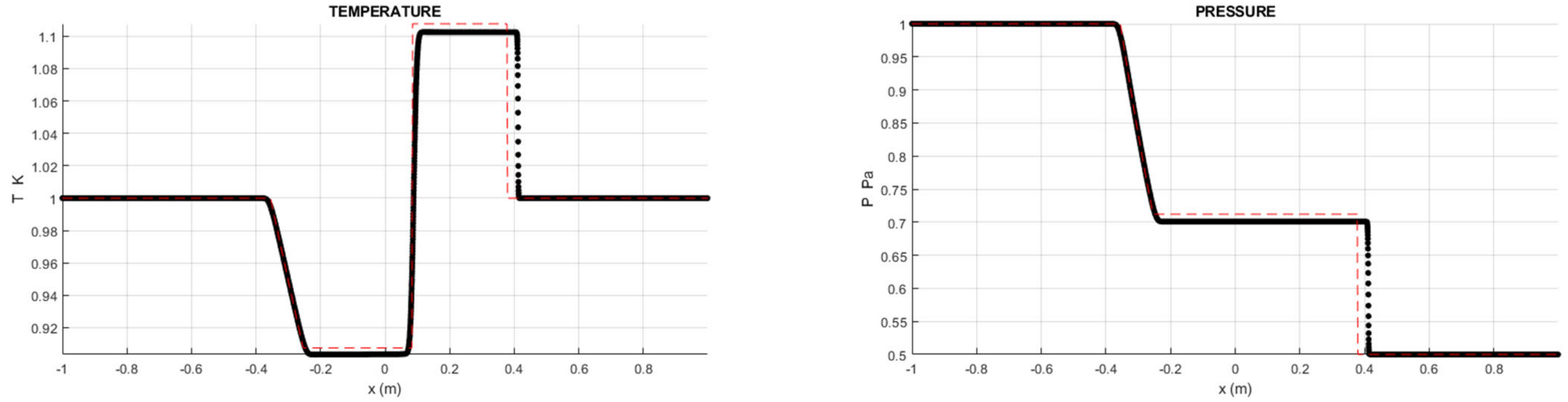

Figure #6 – Temperature & Pressure ($t = 0.3$ s)

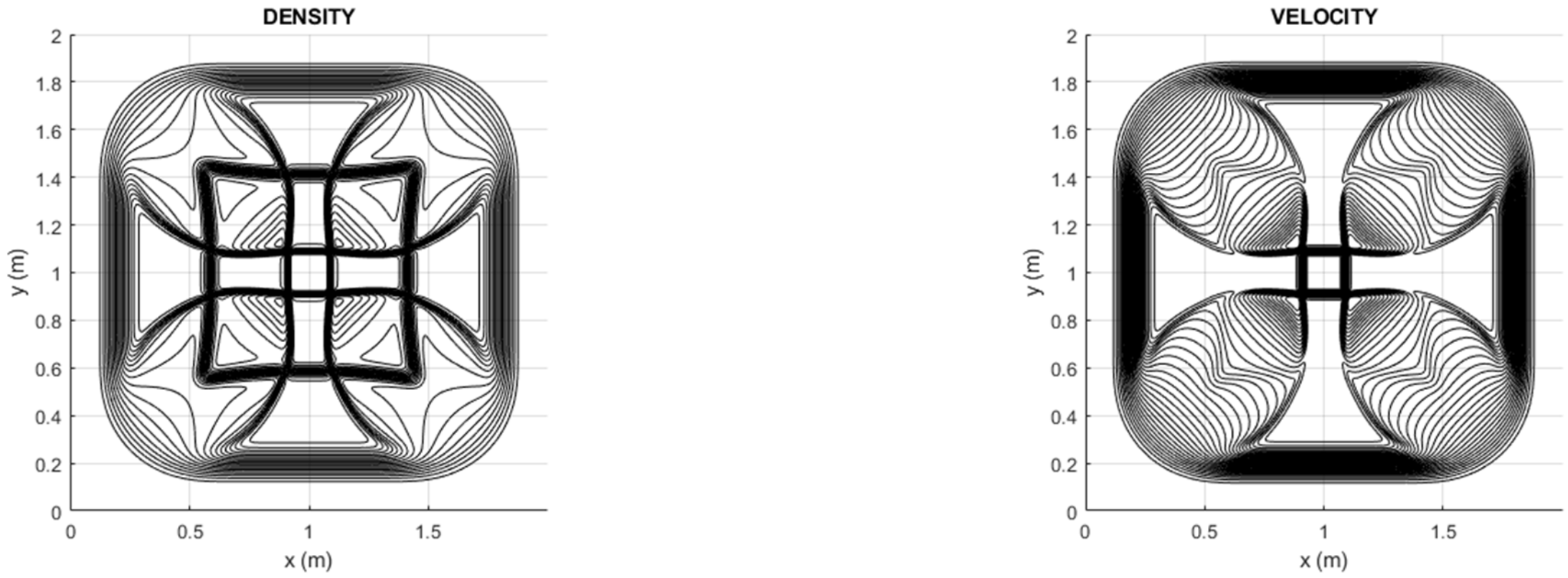

Figure #7 – Density & Velocity ($t = 0.3$ s)

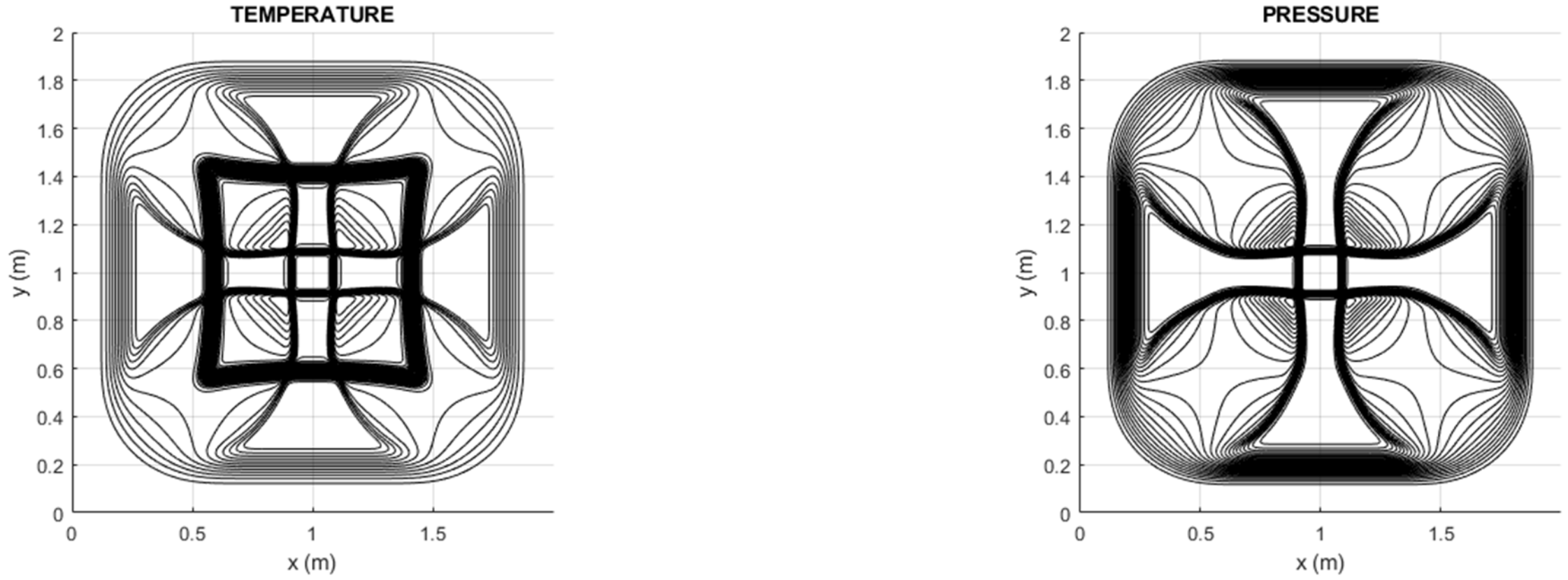

Figure #8 – Temperature & Pressure ($t = 0.3$ s)

The end of this work has been reached. It has already been summarized what has been accomplished in the introduction. What remains is application at the N-S level. Future work may include investigation of (82) for application. Based on the multiscale moment method presented, the incorporation of the stress tensor and heat flux vector in distribution form will need to be investigated.

## APPENDIX

Please see below for a complete list of the Kronecker delta identities used in this work.

$$\delta_{ij} = \begin{matrix} 1 & i = j \\ 0 & i \neq j \end{matrix}$$

Total Kronecker Deltas

$$\delta_{ijkl} = \delta_{ij}\delta_{kl} + \delta_{ik}\delta_{jl} + \delta_{il}\delta_{jk}$$

$$\delta_{ijklmn} = \delta_{ij}\delta_{klmn} + \delta_{ik}\delta_{jlmn} + \delta_{il}\delta_{kjmn} + \delta_{im}\delta_{jkln} + \delta_{in}\delta_{jklm}$$

$$\delta_{ijklmnop} = \delta_{ij}\delta_{klmnop} + \delta_{ik}\delta_{jlmnop} + \delta_{il}\delta_{jkmnop} + \delta_{im}\delta_{jklnop} + \delta_{in}\delta_{jklmop} + \delta_{io}\delta_{jklmnp} + \delta_{ip}\delta_{jklmno}$$

Partial or Starred Kronecker Deltas

$$\delta^*_{ijkl} = \delta_{ik}\delta_{jl} + \delta_{il}\delta_{jk}$$

$$\delta^*_{ijklmn} = \delta_{il}\delta^*_{kjmn} + \delta_{im}\delta^*_{jkln} + \delta_{in}\delta^*_{jklm}$$

$$\delta^{**}_{ijklmn} = \delta_{ij}\delta^*_{klmn} + \delta_{ik}\delta^*_{jlmn} + \delta_{il}\delta^*_{kjmn} + \delta_{im}\delta_{jkln} + \delta_{in}\delta_{jklm}$$

$$\delta^*_{ijklmnop} = \delta_{im}\delta^*_{jklnop} + \delta_{in}\delta^*_{jklmop} + \delta_{io}\delta^*_{jklmnp} + \delta_{ip}\delta^*_{jklmno}$$

Please see below for the Hermite expanded integrals in reference to the required weighting conditions.

Expanded Hermite Integration

The Hermite moments will be abbreviated as:

$$\langle \boldsymbol{H}^{(n)} \boldsymbol{\xi}^m \rangle = \int_{-\infty}^{+\infty} \omega \boldsymbol{H}^{(n)} \boldsymbol{\xi}^m \, \mathrm{d}\boldsymbol{\xi} = \sum_{\alpha=1}^{Q} w_\alpha \boldsymbol{H}_\alpha^{(n)} \boldsymbol{\xi}_\alpha^m$$

The expansion numbers and moments of the Maxwellian are given below:

$n = 0$

$$\mathbb{E}[\Xi^0] = \boldsymbol{A}^0 \cdot \langle \boldsymbol{H}^0 \rangle$$

$n = 1$

$$\mathbb{E}[\Xi^0] = \boldsymbol{A}^0 \cdot \langle \boldsymbol{H}^0 \rangle + \boldsymbol{A}^1 \cdot \langle \boldsymbol{H}^1 \rangle$$

$$\mathbb{E}[\Xi_i] = \boldsymbol{A}^0 \cdot \langle \boldsymbol{H}^0 \Xi_i \rangle + \boldsymbol{A}^1 \cdot \langle \boldsymbol{H}^1 \Xi_i \rangle$$

$\underline{n = 2}$

$$\mathbb{E}[\Xi^0] = \boldsymbol{A}^0 \cdot \langle \boldsymbol{H}^0 \rangle + \boldsymbol{A}^1 \cdot \langle \boldsymbol{H}^1 \rangle + \frac{1}{2}\boldsymbol{A}^2 \cdot \langle \boldsymbol{H}^2 \rangle$$

$$\mathbb{E}[\Xi_i] = \boldsymbol{A}^0 \cdot \langle \boldsymbol{H}^0 \Xi_i \rangle + \boldsymbol{A}^1 \cdot \langle \boldsymbol{H}^1 \Xi_i \rangle + \frac{1}{2}\boldsymbol{A}^2 \cdot \langle \boldsymbol{H}^2 \Xi_i \rangle$$

$$\mathbb{E}[\Xi_{ij}] = \boldsymbol{A}^0 \cdot \langle \boldsymbol{H}^0 \Xi_{ij} \rangle + \boldsymbol{A}^1 \cdot \langle \boldsymbol{H}^1 \Xi_{ij} \rangle + \frac{1}{2}\boldsymbol{A}^2 \cdot \langle \boldsymbol{H}^2 \Xi_{ij} \rangle$$

$\underline{n = 3}$

$$\mathbb{E}[\Xi^0] = \boldsymbol{A}^0 \cdot \langle \boldsymbol{H}^0 \rangle + \boldsymbol{A}^1 \cdot \langle \boldsymbol{H}^1 \rangle + \frac{1}{2}\boldsymbol{A}^2 \cdot \langle \boldsymbol{H}^2 \rangle + \frac{1}{6}\boldsymbol{A}^3 \cdot \langle \boldsymbol{H}^3 \rangle$$

$$\mathbb{E}[\Xi_i] = \boldsymbol{A}^0 \cdot \langle \boldsymbol{H}^0 \Xi_i \rangle + \boldsymbol{A}^1 \cdot \langle \boldsymbol{H}^1 \Xi_i \rangle + \frac{1}{2}\boldsymbol{A}^2 \cdot \langle \boldsymbol{H}^2 \Xi_i \rangle + \frac{1}{6}\boldsymbol{A}^3 \cdot \langle \boldsymbol{H}^3 \Xi_i \rangle$$

$$\mathbb{E}[\Xi_{ij}] = \boldsymbol{A}^0 \cdot \langle \boldsymbol{H}^0 \Xi_{ij} \rangle + \boldsymbol{A}^1 \cdot \langle \boldsymbol{H}^1 \Xi_{ij} \rangle + \frac{1}{2}\boldsymbol{A}^2 \cdot \langle \boldsymbol{H}^2 \Xi_{ij} \rangle + \frac{1}{6}\boldsymbol{A}^3 \cdot \langle \boldsymbol{H}^3 \Xi_{ij} \rangle$$

$$\mathbb{E}[\Xi_{ijk}] = \boldsymbol{A}^0 \cdot \langle \boldsymbol{H}^0 \Xi_{ijk} \rangle + \boldsymbol{A}^1 \cdot \langle \boldsymbol{H}^1 \Xi_{ijk} \rangle + \frac{1}{2}\boldsymbol{A}^2 \cdot \langle \boldsymbol{H}^2 \Xi_{ijk} \rangle + \frac{1}{6}\boldsymbol{A}^3 \cdot \langle \boldsymbol{H}^3 \Xi_{ijk} \rangle$$

$\underline{n = 4}$

$$\mathbb{E}[\Xi^0] = \boldsymbol{A}^0 \cdot \langle \boldsymbol{H}^0 \rangle + \boldsymbol{A}^1 \cdot \langle \boldsymbol{H}^1 \rangle + \frac{1}{2}\boldsymbol{A}^2 \cdot \langle \boldsymbol{H}^2 \rangle + \frac{1}{6}\boldsymbol{A}^3 \cdot \langle \boldsymbol{H}^3 \rangle + \frac{1}{24}\boldsymbol{A}^4 \cdot \langle \boldsymbol{H}^4 \rangle$$

$$\mathbb{E}[\Xi_i] = \boldsymbol{A}^0 \cdot \langle \boldsymbol{H}^0 \Xi_i \rangle + \boldsymbol{A}^1 \cdot \langle \boldsymbol{H}^1 \Xi_i \rangle + \frac{1}{2}\boldsymbol{A}^2 \cdot \langle \boldsymbol{H}^2 \Xi_i \rangle + \frac{1}{6}\boldsymbol{A}^3 \cdot \langle \boldsymbol{H}^3 \Xi_i \rangle + \frac{1}{24}\boldsymbol{A}^4 \cdot \langle \boldsymbol{H}^4 \Xi_i \rangle$$

$$\mathbb{E}[\Xi_{ij}] = \boldsymbol{A}^0 \cdot \langle \boldsymbol{H}^0 \Xi_{ij} \rangle + \boldsymbol{A}^1 \cdot \langle \boldsymbol{H}^1 \Xi_{ij} \rangle + \frac{1}{2}\boldsymbol{A}^2 \cdot \langle \boldsymbol{H}^2 \Xi_{ij} \rangle + \frac{1}{6}\boldsymbol{A}^3 \cdot \langle \boldsymbol{H}^3 \Xi_{ij} \rangle + \frac{1}{24}\boldsymbol{A}^4 \cdot \langle \boldsymbol{H}^4 \Xi_{ij} \rangle$$

$$\mathbb{E}[\Xi_{ijk}] = \boldsymbol{A}^0 \cdot \langle \boldsymbol{H}^0 \Xi_{ijk} \rangle + \boldsymbol{A}^1 \cdot \langle \boldsymbol{H}^1 \Xi_{ijk} \rangle + \frac{1}{2}\boldsymbol{A}^2 \cdot \langle \boldsymbol{H}^2 \Xi_{ijk} \rangle + \frac{1}{6}\boldsymbol{A}^3 \cdot \langle \boldsymbol{H}^3 \Xi_{ijk} \rangle + \frac{1}{24}\boldsymbol{A}^4 \cdot \langle \boldsymbol{H}^4 \Xi_{ijk} \rangle$$

$$\mathbb{E}[\Xi_{ijkl}] = \boldsymbol{A}^0 \cdot \langle \boldsymbol{H}^0 \Xi_{ijkl} \rangle + \boldsymbol{A}^1 \cdot \langle \boldsymbol{H}^1 \Xi_{ijkl} \rangle + \frac{1}{2}\boldsymbol{A}^2 \cdot \langle \boldsymbol{H}^2 \Xi_{ijkl} \rangle + \frac{1}{6}\boldsymbol{A}^3 \cdot \langle \boldsymbol{H}^3 \Xi_{ijkl} \rangle + \frac{1}{24}\boldsymbol{A}^4 \cdot \langle \boldsymbol{H}^4 \Xi_{ijkl} \rangle$$